\documentclass[11pt,a4paper]{article}
\usepackage{jheppub}
\usepackage{amsmath,amssymb}
\usepackage{graphicx}
\usepackage{dcolumn}
\usepackage{bm}
\usepackage{slashed}
\usepackage{setspace} 
\usepackage{ulem}
\usepackage{braket}
\usepackage{color}
\usepackage{subfigure}
\usepackage{float}
\usepackage{amsfonts}

\usepackage{caption}
\usepackage{enumitem}
\begin{document}

\title{{Lepton flavor violation in Higgs boson decays at the HL-LHC}}

\author[a,b]{M. A. Arroyo-Ure\~na }
\author[c]{E. A. Herrera-Chac\'on}
\author[a,b]{Iran Melendez-Hernández}
\author[a,b]{S. Rosado-Navarro}

\affiliation[a]{Facultad de Ciencias F\'isico-Matem\'aticas, Benem\'erita Universidad Aut\'onoma de Puebla, C.P. 72570, Puebla, M\'exico}
\affiliation[b]{Centro Interdisciplinario de Investigaci\'on y Ense\~nanza de la Ciencia (CIIEC), Benem\'erita Universidad Aut\'onoma de Puebla, C.P. 72570, Puebla, M\'exico}
\affiliation[c]{Institute for Particle Physics Phenomenology, Durham University, South Road, Durham, DH1 3LE}

\emailAdd{marco.arroyo@fcfm.buap.mx}
\emailAdd{edwin.a.herrera-chacon@durham.ac.uk}
\emailAdd{iran.melendezh@alumno.buap.mx}
\emailAdd{sebastian.rosado@protonmail.com}

\maketitle
\begin{abstract}
	
We explore the discovery potential for lepton flavor-violating Higgs boson decays, $h \to \ell_i \ell_j$ ($\ell_i \neq \ell_j$; $\ell = e, \mu, \tau$), in proton-proton collisions. The analysis is performed within a Froggatt-Nielsen framework, which offers a well-motivated theoretical mechanism for such processes. By performing a systematic scan over the phenomenologically viable parameter space, we identify regions that can yield observable signatures at the High-Luminosity LHC (HL-LHC). For an integrated luminosity of $\mathcal{L}_{\rm int} \gtrsim 1000~\text{fb}^{-1}$, we project that the channels $h \to e\mu$ and $h \to \tau\mu$ (with $\tau^{\pm} \to \pi^{\pm} \nu_{\tau}$) could reach the $5\sigma$ discovery threshold, while $h \to \tau e$ remains inaccessible even at $3000~\text{fb}^{-1}$. Our results highlight the unique capability of the HL-LHC to probe new physics through searches for rare Higgs decays.
\end{abstract}

\keywords{Lepton Flavor Violation, Flavon Model, Higgs Decay, HL-LHC}

\section{Introduction} 
Lepton flavor violation (LFV) refers to transitions between the charged lepton families—$e$, $\mu$, and $\tau$—that do not conserve lepton family number. In the Standard Model (SM) with massless neutrinos, these individual lepton numbers are strictly conserved. While the introduction of non-zero neutrino masses allows for such processes, their expected rates are highly suppressed by factors of $m^2_{\nu}/m^2_W$ \cite{Raidal:2008jk}, rendering them exceptionally sensitive probes of Beyond the SM (BSM) physics.

The phenomenon of neutrino oscillations, a direct quantum mechanical consequence of non-zero neutrino masses and mixing, provides an established manifestation of LFV in the neutral sector. This has been convincingly demonstrated by experiments utilizing solar, atmospheric, reactor, and accelerator neutrinos \cite{SNO:2002tuh, KamLAND:2004mhv, Super-Kamiokande:2004orf, K2K:2006yov}, confirming the theoretical foundations laid out in \cite{Pontecorvo:1957cp, Maki:1962mu}.

In contrast, the observation of charged Lepton Flavor-Violating (cLFV) decays would represent an unambiguous signature of new physics. To date, no evidence has been found for processes such as the lepton decays $\tau^{-} \rightarrow e^{-} e^{-} e^{+}$, $\tau^{-} \rightarrow \mu^{-} \mu^{-} \mu^{+}$ \cite{Hayasaka:2010np}, and $\mu^{-} \rightarrow e^{-} e^{-} e^{+}$ \cite{SINDRUM:1987nra}, or the radiative decays $\mu \rightarrow e \gamma$ \cite{MEG:2016leq}, $\tau \rightarrow \mu \gamma$ \cite{Belle:2021ysv}, and $\tau \rightarrow e \gamma$ \cite{BaBar:2009hkt}. However, these searches have resulted in very stringent upper limits on the corresponding branching ratios \cite{Crivellin_2014, Asadi_2026}.

A particularly compelling LFV process is the Higgs boson decay $h \to \tau\mu$, initially investigated in \cite{Pilaftsis:1992st, Korner:1992zk, Diaz-Cruz:1999sns} and analyzed its potential detectability in \cite{Han:2000jz, Assamagan:2002kf}. This decay has been extensively explored within several SM extensions, including models with massive neutrinos, extended scalar, and supersymmetric models \cite{Arganda:2004bz, Lami:2016mjf, Barman_2023, Arganda_2015, Crivellin:2015mga, Arganda_2016, Arganda_2017, Marcano_2020, Arroyo-Urena:2023vfh, Urena:2021xtw, Arroyo-Urena:2020mgg}.
Less studied have been the processes $h\to \tau e$ and $h\to e\mu$, which have a special interest in this work.

Current experimental upper limits on these channels, as reported by the CMS and ATLAS collaborations \cite{ATLAS:2023mvd, CMS:2021rsq}, exhibit  the values $\mathcal{BR}(h\to\tau\mu)<1.5\times 10^{-3}$, $\mathcal{BR}(h\to\tau e)<2\times 10^{-3}$, and $\mathcal{BR}(h\to e\mu)<4.4\times 10^{-5}$. Although these limits are not yet highly restrictive\footnote{Although, as will be elucidated in Sec. \ref{sec:SecII}, the $h \to \tau\mu$ decay channel is stringently constrained by the current experimental upper limit on the branching ratio $\mathcal{BR}(h \to \tau\mu)$.}, they establish a baseline for future searches for the $h\to\ell_i\ell_j$ decays with integrated luminosities larger than the one reached by the LHC ($300$ fb$^{-1}$). This could be achieved at the HL-LHC \cite{Cerri:2018ypt}, which will represent a new stage of the LHC expected to begin around 2026 with a center-of-mass energy of 14 TeV. The upgrade aims at increasing the integrated luminosity by a factor of ten ($3000$ fb$^{-1}$, around the year 2035) with respect to the final stage of the LHC. 

This work investigates the decays $h \to \ell_i\ell_j$ within a theoretical framework that extends the SM scalar sector with a complex singlet and incorporates the Froggatt-Nielsen mechanism \cite{Froggatt:1978nt}, which provides a potential explanation for the hierarchical structure of fermion masses and mixings. We explore realistic benchmark points (BP) in the model's parameter space which yield different production cross sections and, consequently, different projected significances, making them prime targets for investigation at the HL-LHC.

This work is structured as follows. In Sec.~\ref{sec:SecII}, we conduct a comprehensive review of the Froggatt-Nielsen Singlet Model (FNSM). Experimental and theoretical constraints on the model parameter space are also included. Section~\ref{sec:SecIII} focuses on taking advantage of the insights gained from the previous section, performing a computational analysis of the proposed signal and its SM background processes. Finally, the conclusions are presented in Sec.~\ref{sec:SecIV}.

\section{Theoretical framework}\label{sec:SecII}
In this section we present the relevant theoretical aspects of the FNSM. In Refs.~\cite{Arroyo-Urena:2025zvg,Arroyo-Urena:2022oft, Arroyo-Urena:2018mvl, Bauer:2016rxs, Abbas:2024dfh, Arroyo-Urena:2019fyd,Greljo:2024evt, Loisa:2024xuk, Kikuchi:2023fpl, Asadi:2023ucx} a deep theoretical analysis  of the model and a study on the model parameter space are also reported.

\subsection{Scalar sector}
The scalar sector is extended by adding a complex singlet scalar field, $S_F$, to the SM. In the unitary gauge, the SM Higgs doublet $\Phi$ and the new singlet $S_F$ are expressed as:
\[
\Phi=\left(\begin{array}{c}
0\\
\frac{v+\phi^{0}}{\sqrt{2}}
\end{array}\right),\quad S_{F}=\left(\frac{v_{s}+S_{R}+iS_{I}}{\sqrt{2}}\right),
\]
where $v$ and $v_s$ denote the vacuum expectation values (VEVs) of the Higgs doublet and the complex singlet, respectively. The scalar potential is required to be invariant under a flavor-dependent $U(1)_F$ symmetry, ---following the original Froggatt-Nielsen construction, this symmetry is taken as global---under which the fields transform as $\Phi\to\Phi$ and $S_F\to e^{i\alpha} S_F$. In the general FNSM, the scalar potential allows for a complex VEV for the singlet $\braket{S_F}_0=\frac{v_s}{\sqrt{2}}e^{i\xi}$. However, for this work, we restrict our analysis to the CP-conserving scenario by setting the phase $\xi=0$. 

Therefore, the CP-conserving scalar potential is then given by:
\begin{equation}\label{potential}
V_0 = -\frac{1}{2}m_1^2\Phi^{\dagger}\Phi - \frac{1}{2}m_2^2 S_F^* S_F + \frac{1}{2}\lambda_1(\Phi^{\dagger}\Phi)^2 + \lambda_2(S_F^* S_F)^2 + \lambda_3(\Phi^{\dagger}\Phi)(S_F^* S_F).
\end{equation}
After spontaneous symmetry breaking (SSB), driven by the VEVs of $\Phi$ and $S_F$, a massless Goldstone boson emerges. To generate a mass for this state, we introduce a soft $U(1)_F$-breaking term to the scalar potential in Eq.~\eqref{potential}:
\begin{equation}
V{\text{soft}} = -\frac{1}{2}m_3^2(S_F^2 + S_F^{*2}).
\end{equation}
The complete scalar potential of the model is therefore
\begin{equation}
V_{\text{FNSM}} = V_0 + V_{\text{soft}}.
\end{equation}
Following SSB and the application of the minimization conditions to the potential  $V_{\rm FNSM}$ are performed, the mixing between the $\phi^0$ and $S_R$ fields, induced by the $\lambda_3$ parameter, leads to the following relations for the mass parameters:
\begin{align}
    m_1^2&=v^2\lambda_1+v_s^2\lambda_3,\\
    m_2^2&=-2m_3^2+2v_s^2\lambda_2+v^2\lambda_3.
\end{align}
These parameters define the scalar mass matrix. Furthermore, the soft $U(1)_F$-breaking term $V_{\rm soft}$ explicitly generates a mass for the pseudo-scalar Flavon field $S_I$.

Since all parameters in the potential \eqref{potential} are real, the resulting mass matrix is block-diagonal, and the CP-even and CP-odd scalar sectors do not mix. In the $(\phi_0,\,S_R)$ basis, the CP-even mass matrix is given by
\[
M_{S}^{2}=\left(\begin{array}{cc}
\lambda_{1}v^{2} & \lambda_{3}vv_{s}\\
\lambda_{3}vv_{s} & 2\lambda_{2}v_{s}^{2}
\end{array}\right).
\]
The mass eigenstates are obtained through the standard $2\times 2$ rotation: 

\begin{equation}
\left(\begin{array}{c}
\phi^{0}\\
S_{R}
\end{array}\right)=\left(\begin{array}{cc}
\cos\alpha & \sin\alpha\\
-\sin\alpha & \cos\alpha
\end{array}\right)\left(\begin{array}{c}
h\\
H_{F}
\end{array}\right),\label{h-HF-mixing}
\end{equation}
where we identify $h$ as the SM-like Higgs boson and $H_F$ is the CP-even Flavon. The CP-odd Flavon is associated with the imaginary part of the complex singlet: $S_I\equiv A_F$ with mass $M_{A_F}^2=2m_3^2$. The physical masses $M_{\phi}$ ($\phi=h,\,H_F,\,A_F$) are related to the parameters of the scalar potential in Eq.\eqref{potential} as follows:
\begin{align}\label{eq:lambdas}
    \lambda_1&=\frac{(\cos\alpha\,M_h)^2+(\sin\alpha\,M_{H_F})^2}{v^2},\nonumber\\
    \lambda_2&=\frac{M_{A_F}^2+(\cos\alpha\,M_{H_F})^2+(\sin\alpha\,M_h)^2}{2v_s^2},\\
    \lambda_3&=\frac{\cos\alpha\sin\alpha}{vv_s}(M_{H_F}^2-M_h^2).\nonumber
\end{align}

\subsection{Yukawa Lagrangian}
 The Yukawa Lagrangian, invariant under the $U(1)_F$ symmetry, is given by \cite{Froggatt:1978nt}:
\begin{eqnarray} \label{eq:Yuklag}
    \mathcal{L}_ Y &=& \rho^d_{ ij } \left( \frac{S_F}{\Lambda} 
		\right)^{q_{ ij }^d}  \bar{Q}_i d_{R_j}   \Phi 
		+ \rho^u_{ ij } \left(\frac{S_F}{\Lambda }\right)^{q_{ij }^u}\bar{Q}_i u_{R_j} 
		\tilde\Phi \nonumber \nonumber\\
        &+&\rho^\ell_{ij}\left(\frac{S_F}{\Lambda}\right)^{q_{ij}^\ell}
		\bar{L}_i \ell_{R_j} \Phi+\rho^\nu_{ij}\left(\frac{S_F}{\Lambda}\right)^{q_{ij}^{\nu}}
		\bar{L}_i \nu_{R_j} \tilde{\Phi}  + \rm h.c., 
 \end{eqnarray} 
where $\rho^{f}_{ij}$ ($f=u,d,\ell,\nu$) are $\mathcal{O}(1)$ dimensionless couplings, $q_{ij}^f$ are the flavor charges chosen to reproduce the observed fermion mass hierarchy, and $\Lambda$ is the ultraviolet cutoff scale. 

After SSB of the $U(1)_{F}$ and electroweak symmetries, the Lagrangian in Eq.~\eqref{eq:Yuklag} generates the effective Yukawa couplings. By considering the unitary gauge and making the expansion of the neutral component of the heavy Flavon $S_F$ around its VEV $v_s$, one obtains:
\begin{equation} \label{expansion}
		\Bigg(\frac{S_F}{\Lambda}\Bigg)^{q_{ ij }} 
		\simeq \left(\frac{v_s}{ \sqrt 2\Lambda}  \right)^{q_{ ij }} \left[1+q_{ ij }\left(\frac{S_R+iS_I}{v_s}\right)\right],
\end{equation}
From Eqs. \eqref{eq:Yuklag}, \eqref{expansion} and after replacing the mass eigenstates, the Yukawa Lagrangian reads:
\begin{eqnarray} \label{Yukalagrangian} 
		\mathcal {L}  _Y &=& \frac 1  v [\bar{U}  M^u U+\bar{D}  M^d D+\bar{L} M^ \ell L](\cos\alpha h+\sin\alpha H_F) \nonumber\\
		&+&\frac{v }{ \sqrt 2 v_s } [\bar{U}_i\tilde Z_{ij}^u U_j+\bar{D}_i\tilde Z_{ij}^d D_j+\bar{L}_i\tilde Z_{ij}^ \ell  L_j]\nonumber\\&\times&
		(-\sin\alpha h+\cos\alpha H_F+iA_F)+ \rm h.c.,  
	\end{eqnarray}
where the Higgs-Flavon couplings are encapsulated in the $\tilde{Z}_{ij}^f=U_L^f Z_{ij}^f U^{f\dagger}_L$ matrix elements and $M^f$ corresponds to the diagonal fermion mass matrix. 
The couplings $\tilde{Z}_{ij}^f$ can be written as follows
\begin{equation}\label{eq:Zij}
    \tilde{Z}_{ij}^f=\tilde{\rho}_{ij}^f\Bigg(\frac{v_s}{\sqrt{2}\Lambda}\Bigg)^{q_{ij}^f},
\end{equation}
where $\tilde{\rho}_{ij}^f$ encompasses the diagonalization effects of $Z_{ij}^f$ into $\rho_{ij}^f$. This means that, in general, $\tilde{\rho}_{ij}^f$ is different from $\mathcal{O}(1)$.
Equation~\eqref{eq:Zij} remains nondiagonal even after the process of diagonalization of the mass matrices, as a consequence, the FNSM allows for flavor-changing neutral currents. This Froggatt-Nielsen mechanism, first applied to generate flavor-changing Higgs couplings in~\cite{Dery_2014}, provides a natural explanation for the observed fermion mass hierarchy while allowing for potentially observable LFV signals.
The exponents $q_{ij}^f$ are determined by the flavor charges of the fermions and the Higgs doublet. For the quark sector, they are given by~\cite{Bauer:2016rxs}: 
\begin{eqnarray}\label{chargesQ}
    q_{ij}^d&=&q_{Q_i}-q_{d_j}-q_H,\nonumber \\
    q_{ij}^u&=&q_{Q_i}-q_{u_j}+q_H,
\end{eqnarray}
where $q_{u_i}=q_{u,\,c,\,t}$ and $q_{d_i}=q_{d,\,s,\,b}$ are the flavor charges of the right-handed quark singlets, $q_{Q_i}$ are the charges of the quark doublets, and $q_H$ is the Higgs doublet charge. To reproduce the measured quark masses, the charges are fixed to $q_S=+1$, $q_H=0$ and

\begin{center}
$\left(\begin{array}{ccc}
q_{Q_{1}} & q_{Q_{2}} & q_{Q_{3}}\\
q_{u} & q_{c} & q_{t}\\
q_{d} & q_{s} & q_{b}\label{chargesQmatrix}
\end{array}\right)=\left(\begin{array}{ccc}
3 & 2 & 0\\
-5 & -2 & 0\\
-4 & -3 & -3
\end{array}\right)$
\par\end{center}
The analogous expressions for the lepton sector in Eq.~\eqref{eq:Zij} are given by

\begin{eqnarray}\label{chargesL}
    q_{ij}^\ell&=&q_{L_i}-q_{\ell_j}-q_H,\nonumber \\
    q_{ij}^\nu&=&q_{L_i}-q_{\nu_j}+q_H,
\end{eqnarray}
where $q_{\nu_j}=q_{\nu_e},\,q_{\nu_\mu},\,q_{\nu_\tau}$ and $q_{\ell_j}=q_e,\,q_\mu,\,q_\tau$ are the flavor charges of the right-handed neutrino and charged lepton singlets, respectively, and $q_{L_i}$ are the charges of the lepton doublets. The specific charge assignments chosen to reproduce the observed lepton masses and mixing pattern are:
\begin{center}
$\left(\begin{array}{ccc}\label{chargesLmatrix}
q_{L_{1}} & q_{L_{2}} & q_{L_{3}}\\
q_{\nu_{e}} & q_{\nu_{\mu}} & q_{\nu_{\tau}}\\
q_{e} & q_{\mu} & q_{\tau}
\end{array}\right)=\left(\begin{array}{ccc}
1 & 0 & 0\\
-24 & -21 & -20\\
-8 & -5 & -3
\end{array}\right)$
\par\end{center}

With the parametrization $\epsilon=v_s/\sqrt{2}\Lambda$, the fermion mass matrices exhibit the following hierarchy:
\begin{eqnarray}\label{fermion_masses}
    \frac{m_b}{m_t}&\approx& \epsilon^3,\;\frac{m_c}{m_t}\approx \epsilon^4,\;\frac{m_s}{m_t}\approx \epsilon^5,\;\frac{m_d}{m_t}\approx \epsilon^7,\;\frac{m_u}{m_t}\approx \epsilon^8\nonumber\\ \frac{m_{\tau}}{m_t}&\approx& \epsilon^3,\;\frac{m_\mu}{m_t}\approx \epsilon^5,\;\frac{m_e}{m_t}\approx \epsilon^9,\;m_t\approx \frac{v}{\sqrt{2}},
\end{eqnarray}
and the CKM matrix becomes
\begin{center}
$V_{{\rm CKM}}\approx\left(\begin{array}{ccc}
1 & \epsilon & \epsilon^{3}\\
\epsilon & 1 & \epsilon^{2}\\
\epsilon^{3} & \epsilon^{2} & 1
\end{array}\right)$.
\par\end{center}

As far as the $\phi VV$ ($V=W,\,Z$) interactions are concerned, from the kinetic terms of the Higgs doublet and the complex singlet we can extract their couplings. Thus, we present the relevant Feynman rules  in Table~\ref{couplings}.
    \begin{table}[htb!]   
	\begin{centering} 
		\begin{tabular}{ cc} 
			\hline
			Vertex ($\phi XX$) &Coupling constant ($g_{\phi XX} $) \tabularnewline
			\hline
			$ hf_ i \bar{f} _ j  $ & $\frac{ \cos \alpha}{   v} \tilde M _{ ij }^ f -\sin \alpha r_ s \tilde Z ^ f _{ ij }$\tabularnewline
			$ H_ F f_ i \bar{f }_ j  $ & $\frac{ \sin \alpha }{  v} \tilde M _{ ij }^ f +\cos \alpha r_ s \tilde Z ^ {f _{ ij }}$\tabularnewline
			$ A_ F f_ i \bar{f} _ j  $ & $ \, r_ s \tilde Z ^{ f _{ ij }}$\tabularnewline
			$ hZZ $ & $\,\frac{ g M_Z}{ c_W}      \cos\alpha $\tabularnewline
			$ hWW $ & $\,g M_W  \cos\alpha $\tabularnewline
			$ H_ F ZZ $ & $\,g\frac{M_Z}{ c_W}    \sin\alpha $\tabularnewline
			$ H_ F WW $ & $\,g M_W  \sin\alpha $\tabularnewline
			\hline
		\end{tabular}
		\caption{Tree-level couplings of the SM-like Higgs boson $h$ and the Flavons $H_F$ and $A_F$ to fermion and gauge boson pairs in the FNSM. 
Here, $r_s=v/(\sqrt 2 v_s)$.}
		\label{couplings}
		\par\end{centering} 
\end{table} 
\\
It is important to note that the FNSM extends only the scalar sector of the SM, leaving the gauge sector unchanged at tree level. Consequently, FCNC mediated by the $Z$ boson do not arise in this model, avoiding stringent constraints from processes such as $Z\to \ell_i\ell_j$. All LFV effects in the FNSM are generated through the exchange of the flavon fields $H_F$, $A_F$ and the Higgs boson.


\subsection{Model parameter space}
The free model parameters that directly influence our phenomenological predictions are the scalar mixing angle \(\alpha\), the vacuum expectation value of the complex singlet \(v_s\), and the off-diagonal flavor-changing parameters \(\tilde{Z}_{e\mu}\) and \(\tilde{Z}_{\tau\mu}\).
While theoretical bounds derived from vacuum stability, perturbativity, and unitarity of the scalar potential impose a mild lower limit of \(v_s \gtrsim 50\ \text{GeV}\)~\cite{Arroyo-Urena:2025zvg}, experimental constraints prove significantly more restrictive when confronting the model parameter space. Our analysis incorporates relevant experimental data, including Higgs boson measurements from the LHC (signal strength modifiers $\mathcal{R}_{X}$) and the current upper limits on the branching ratios for the processes \(\mathcal{BR}(\mu \to e \gamma)<4.2\times 10^{-13}\) \cite{MEG:2016leq}, \(\mathcal{BR}(\tau \to \mu \gamma)<4.2\times 10^{-8}\) \cite{Belle:2021ysv}, \(\mathcal{BR}(\tau \to e \gamma)<3.3\times 10^{-8}\) \cite{BaBar:2009hkt}, $\mathcal{BR}(h\to\tau\mu)<1.5\times 10^{-3}$, $\mathcal{BR}(h\to\tau e)<2\times 10^{-3}$, and $\mathcal{BR}(h\to e\mu)<4.4\times 10^{-5}$ \cite{ATLAS:2023mvd, CMS:2021rsq}, \(\mathcal{BR}(\tau \to 3\mu)<2.1\times 10^{-8}\) \cite{Hayasaka:2010np}, and \(\mathcal{BR}(\mu \to 3e)<1\times 10^{-12}\) \cite{SINDRUM:1987nra}.
To facilitate the evaluation of these experimental constraints, we provide the following analytical expressions for the corresponding branching ratios. A detailed discussion of the theoretical constraints is provided in Sec.~2-C of Ref.~\cite{Arroyo-Urena:2025zvg}, to which we direct the reader. Flavor-changing $Z$ boson decays are induced only at the one-loop level in the FNSM\footnote{A one-loop induced process such as $Z\to \ell_i\ell_j$ is estimated to have a branching ratio of $\mathcal{O}(10^{-9})$ in our framework, which remains well below the current experimental sensitivity.}, yielding branching ratios that are orders of magnitude below current experimental limits \cite{ParticleDataGroup:2024cfk}, and thus do not impose relevant constraints.
\\

\subsubsection{Signal strength modifiers $\mathcal{R}_{X}$}\label{muX}
For a production process via proton-proton collisions $\sigma(pp\to H_i)$ and a decay $H_i\to X$, the signal strength is defined as follows \cite{Cepeda:2019klc}:

\begin{equation}\label{EqRX}
	\mathcal{R}_{X}=\frac{\sigma(pp\to h)\cdot\mathcal{BR}(h\to X)}{\sigma(pp\to h^{\text{SM}})\cdot\mathcal{BR}(h^{\text{SM}}\to X)},
\end{equation}
where $\sigma(pp\to H_i)$ is the production cross section of $H_i$, with $H_i=h,\,h^{\text{SM}}$; here $h$ is the SM-like Higgs boson coming from an extension of the SM and $h^{\text{SM}}$ is the SM Higgs boson; $\mathcal{BR}(H_i\to X)$ is the branching ratio of the decay $H_i\to X$, with $X=c\bar{c},\,b\bar{b}$,\,$\mu^-\mu^+$,\, $\tau^-\tau^+$, $WW^*$,$ZZ^*$, $\gamma\gamma$, $Z\gamma$.

\subsubsection{$\ell_i\to\ell_j\gamma$}

\begin{figure}[!htb]
	\centering
	\includegraphics[width=14cm]{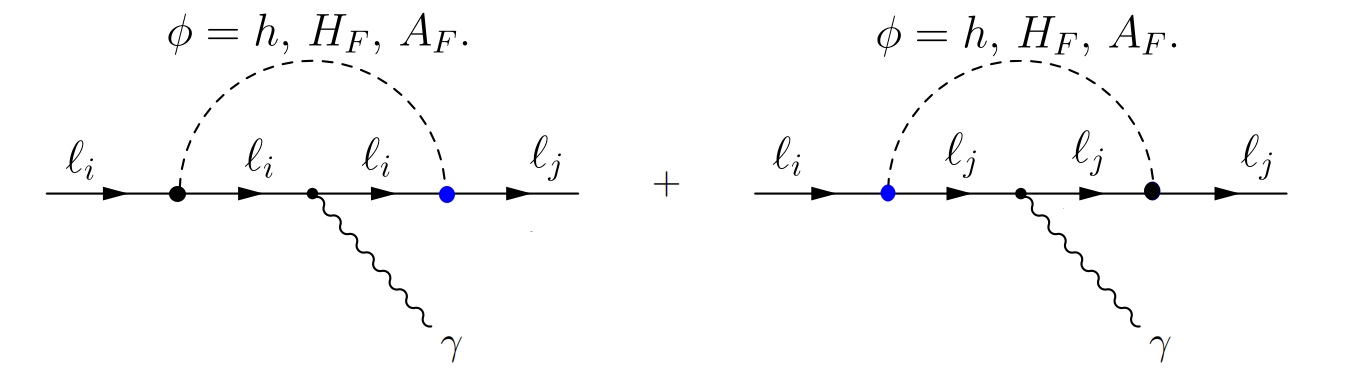}
	\caption{Feynman diagram of the decay $\ell_i\to \ell_j\gamma$. The (pseudo-)scalars $\phi$ induce such a process at one-loop level.}\label{FDliljgamma}
\end{figure}

The FNSM accommodates radiative contributions that induce the decay $\ell_i \to \ell_j \gamma$, whose leading-order diagrams are depicted in Fig. \ref{FDliljgamma}. The associated branching ratio can therefore impose stringent constraints on the parameters governing both the flavor-conserving coupling $g_{\phi\ell_i\ell_i}$ and the flavor-violating coupling $g_{\phi\ell_i\ell_j}$, represented diagrammatically by the black and blue vertices, respectively.

	The effective Lagrangian for the $\ell_i\to\ell_j\gamma$ ($\ell_i\to\ell_j\gamma$ stands for $\tau\to \mu \gamma, \tau\to e \gamma$ and $\mu\to e\gamma$) is given by
	\begin{equation}\label{EffLagrangian}
	\mathcal{L}_{\text{eff}}=C_L Q_{L\gamma} C_R Q_{R\gamma}+h.c.,
	\end{equation}
where the dim-5 electromagnetic penguin operators read
\begin{equation}\label{dim5Op}
Q_{L\gamma,\,R\gamma}=\frac{e}{8\pi^2}(\bar{\ell}_j\sigma^{\alpha\beta}P_{L,\,R}\ell_i)F_{\alpha\beta},
\end{equation}
where $F_{\alpha\beta}$ is the electromagnetic field strength tensor. The Wilson coefficients $C_{L,\,R}$ receive contributions at one-loop level and an important contribution from Barr-Zee two-loops level. For the particular case when $\ell_i=\mu$ and $\ell_j=e$, the one-loop Wilson coefficients $C_{L,\,R}$ simplify as follows \cite{Harnik:2012pb, Blankenburg:2012ex}
\begin{equation}\label{WilsonCoe1loop}
C_L^{1-loop} = \sum_{\phi}\frac{g_{\phi\mu\mu}g_{\phi\mu e}^*}{12m_\phi^2}\Bigg(-4+3\log\frac{m_\phi^2}{m_{\mu}^2}\Bigg),\;\;\;C_R^{1-loop}=\sum_{\phi}\frac{g_{\phi\mu\mu}g_{\phi e\mu}}{12m_\phi^2}\Bigg(-4+3\log\frac{m_\phi^2}{m_{\mu}^2}\Bigg),
\end{equation}
The numerical expression for 2-loop contributions is given by
\begin{equation}\label{WilsonCoe2loop}
C_L^{2-loop}=\sum_{\phi}g_{\phi\mu e}^*(-0.082g_{\phi tt}+0.11)/(m_\phi\text{GeV})^2.
\end{equation}
where $Y_{tt}=\bar{m}_t/v=0.67$, with $\bar{m}_t\simeq 164 \text{GeV}$. The term $g_{\phi P_1P_2}$ stands for the $\phi  P_1P_2$ coupling --given in Table~\ref{couplings}--, where $\phi =h,\,H_F,\,A_F$ and $P_1P_2=\mu\mu,\,\mu e,\,tt$.

The rate for $\mu\to e\gamma$ is
\begin{equation}\label{ratetaumugamma}
\Gamma(\mu\to e\gamma)=\frac{\alpha m_{\mu}^2}{64\pi^4}(|C_L|^2+|C_R|^2).
\end{equation}
To obtain the corresponding width decay of the processes $\tau\to \mu\gamma$ and $\tau\to e\gamma$, the replacements $\mu\to \tau,\, e\to \mu$ for the first decay and $\mu\to \tau$ for the second process from \eqref{WilsonCoe1loop} to \eqref{ratetaumugamma} are required. 
\subsubsection{$h\to \ell_i \ell_j$ decays}\label{hlilj}
\begin{figure}[!htb]
	\centering
	\includegraphics[width=5cm]{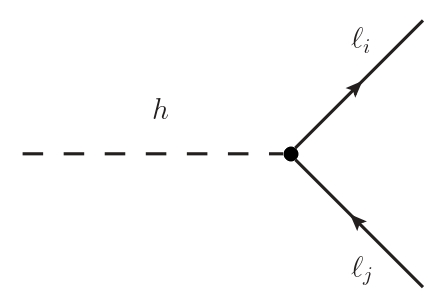}
	\caption{Feynman diagram of the decay $h\to \ell_i \ell_j$. This decay can constrain parameters involved in the $h\ell_i\ell_j$ interaction, indicated by the black dot.}\label{FDhlilj}
\end{figure}

	The LFV processes $h\to \ell_i\ell_j$ ($\ell_{i,\,j}=\ell_{i,\,j}^\mp,\,\ell_{j,\,i}^\pm$) where $\ell_i\ell_j=e\mu,\,e\tau,\,\tau\mu$ arises at tree level in the FNSM. 
The corresponding decay width of the process $h\to f_i \bar{f}_j$ is given by:
\begin{equation}\label{DeacyWid_Higgs-fermions}
\Gamma(h\to \bar{f}_i f_j)=\frac{N_c g^2_{h \bar{f}_i f_j }m_h}{128\pi}\Bigg[ 4-(\sqrt{\tau_{f_i}}+\sqrt{\tau_{f_j}})^2  \Bigg]^{3/2}\sqrt{4-(\sqrt{\tau_{f_i}}-\sqrt{\tau_{f_j}})^2},
\end{equation}
where $g_{h \bar{f}_i f_j}$ is the $h\bar{f}_i f_j$ coupling, $Nc=3\,(1)$ is the color number for quarks (leptons), $m_h$ is the Higgs boson mass and $\tau_{f_{i}}=4m_{f_i}^2/m_h^2$. Note that in general $i\neq j$.
\\
Figure \ref{FDhlilj} presents the Feynman diagram at tree-level that contribute to the process $h\to\ell_i\ell_j$. Given the coupling $g_{h\ell_i\ell_j} = -\frac{\sin\alpha\, v}{\sqrt{2}\,v_s}\tilde{Z}_{ij}$ (for $i\neq j$), upper bounds on $\mathcal{BR}(h \to \ell_i\ell_j)$ can be used to impose direct constraints on the parameters that govern its phenomenological predictions, namely the mixing angle $\alpha$, the flavor-violating matrix element $\tilde{Z}_{ij}$, and the singlet vacuum expectation value $v_s$.

\subsubsection{$\ell_i\to\ell_j\ell_k\bar{\ell}_k$ decays}
\begin{figure}[!htb]
	\centering
	\includegraphics[width=5cm]{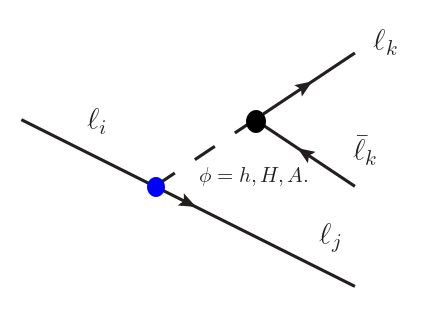}
	\caption{Feynman diagrams that contribute to the $\ell_i\to \ell_j\ell_k\bar{\ell}_k$.}\label{FDliljlklk}
\end{figure}

\begin{figure}[!htb]
  \centering
  \includegraphics[scale=0.2]{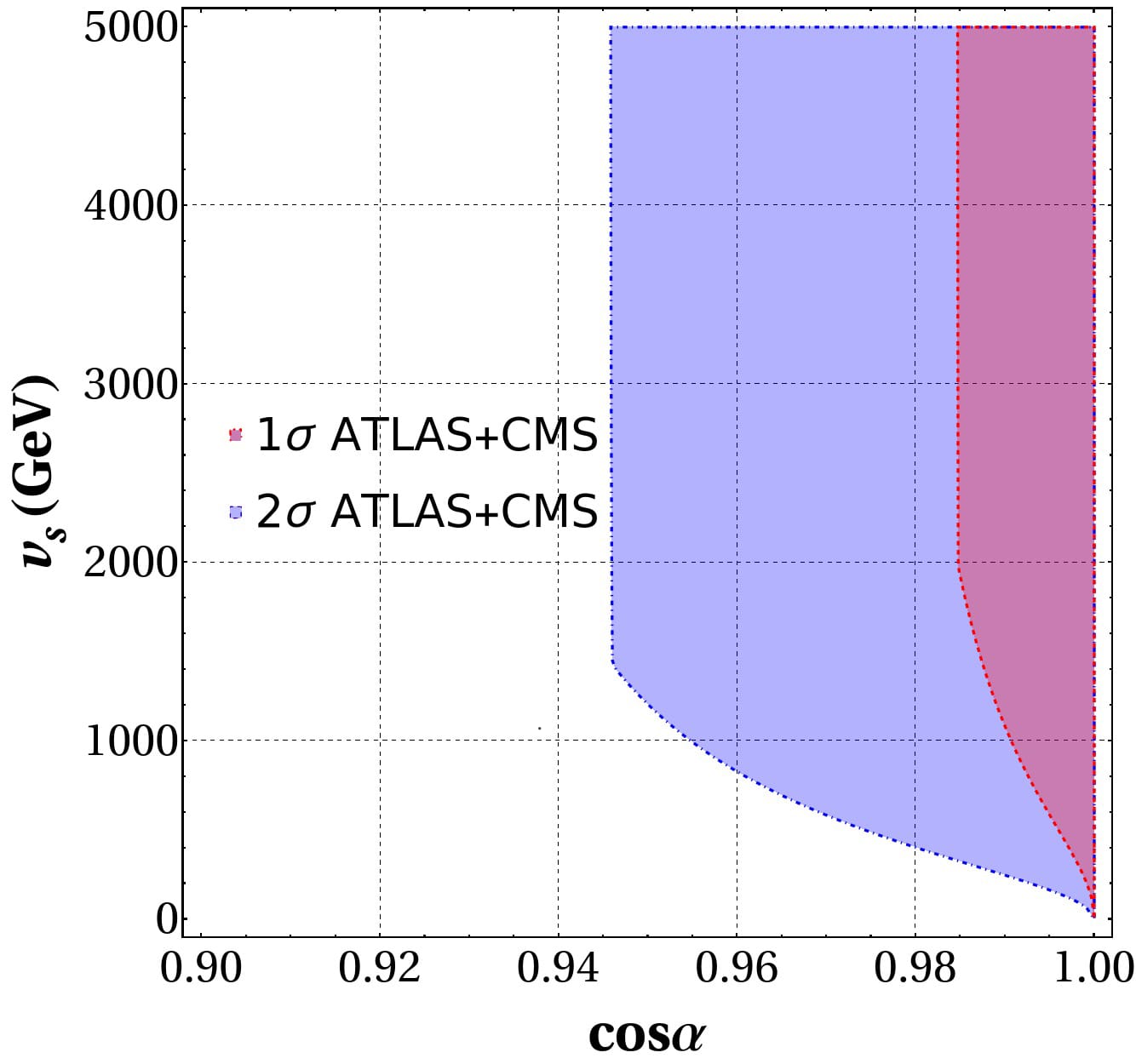}
     \caption{Allowed parameter space by the signal strength modifiers $\mathcal{R}_{X}$ in the $\cos\alpha-v_s$ plane at 1$\sigma$ (blue region) and $2\sigma$ (red area).}\label{signal_strength}
 \end{figure}

Within the FNSM, such decays can proceed at tree level through the exchange of both the SM-like Higgs boson and the flavon fields ($h$, $H_F$, $A_F$), as illustrated in Fig. \ref{FDliljlklk}. Following the convention established in prior analyses, the blue and black points represent the flavor-violating coupling $g_{\phi\ell_i\ell_j}$ and the flavor-conserving coupling $g_{\phi\ell_k\bar{\ell}_k}$, respectively. As delineated in Table \ref{couplings}, these couplings are explicitly parameterized by the mixing angle $\alpha$, the off-diagonal matrix element $\tilde{Z}_{ij}$, and the singlet complex VEV $v_s$. The branching ratios for the processes $\ell_i\to\ell_j\ell_k\bar{\ell}_k$ are sensitive to the product of the relevant flavor-violating and flavor-conserving couplings, $\tilde{Z}_{\ell_i\ell_j}$ and $\tilde{Z}_{\ell_k\ell_k}$, respectively, which can lead to significant phenomenological suppression depending on the specific values of the $\tilde{Z}_{ij}$ matrix elements. 
\\
The corresponding expression for the flavor violating partial width of the  $\tau$ is

\begin{equation}
    \Gamma(\tau\to 3 \mu)\approx \frac{\alpha^2 m^5_\tau} {6(2\pi)^5}\Big|\log\frac{m_\mu^2}{m_\tau^2}-\frac{11}{4}\Big|(|C_L|^2+|C_R|^2), 
\end{equation}
where $C_{L,\,R}$ are given approximately by Eqs. \eqref{WilsonCoe1loop} and \eqref{WilsonCoe2loop}.  
\\
\\
Based on Eq.~\eqref{EqRX}, the parameter region consistent with signal strength modifiers $\mathcal{R}_{X}$—derived from the combined ATLAS and CMS analyses \cite{ATLAS:2021vrm, Cepeda:2019klc}—is mapped onto the $\cos\alpha$–$v_s$ plane, as illustrated in Fig.~\ref{signal_strength}. 
As anticipated, the structure of the Higgs–fermion couplings,
\[
g_{hf_i\bar{f}_j} = \cos\alpha \, \frac{m_f}{v} - \sin\alpha \, \frac{v\,\tilde{Z}_{ij}}{\sqrt{2}v_s},
\]
strongly favors $\cos\alpha \approx 1$, thereby preserving agreement with the predictions of the SM.

\begin{figure}[!htb]
  \centering
  \subfigure[]{\includegraphics[scale=0.17]{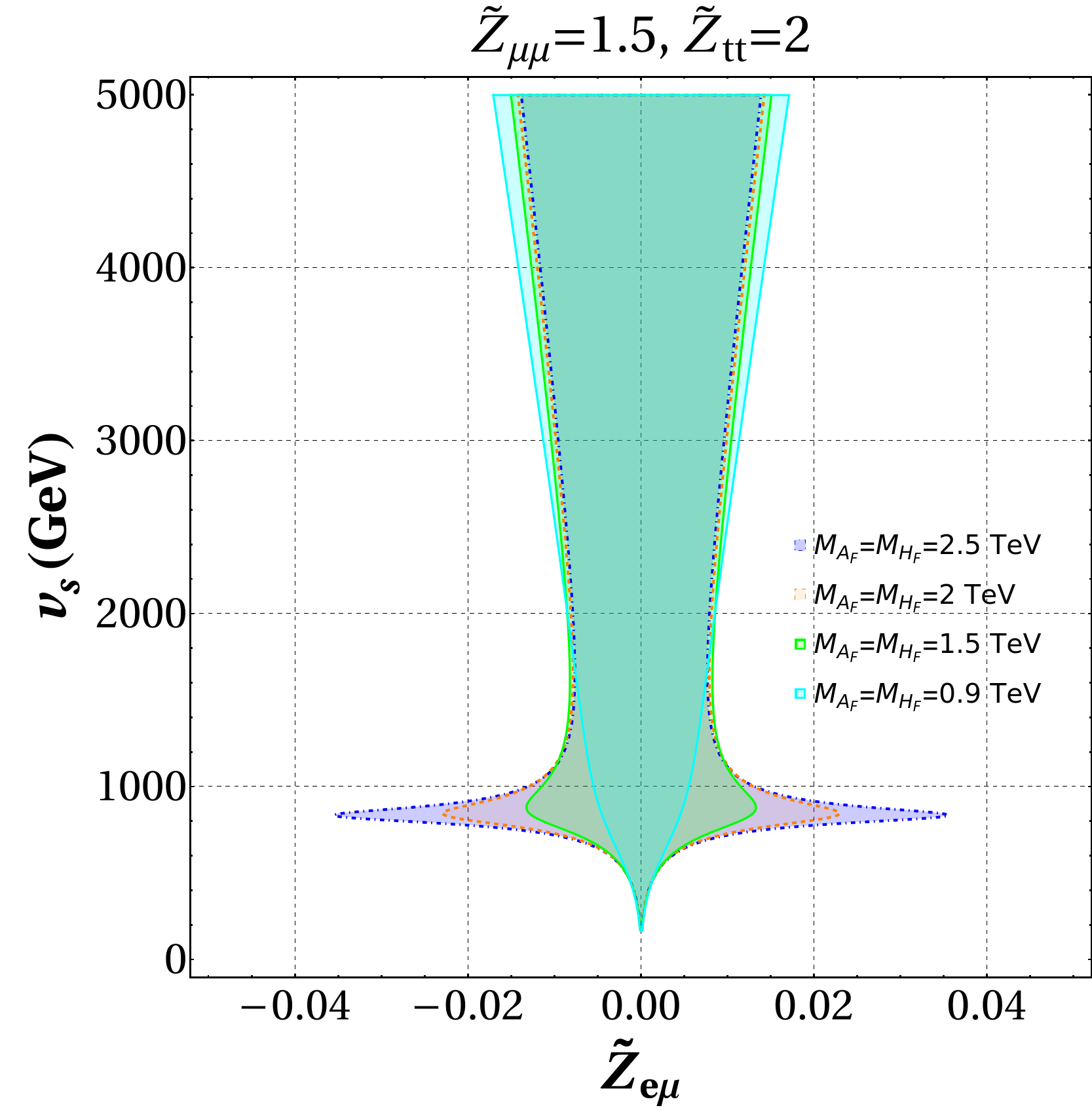}}
    \subfigure[]{\includegraphics[scale=0.17]{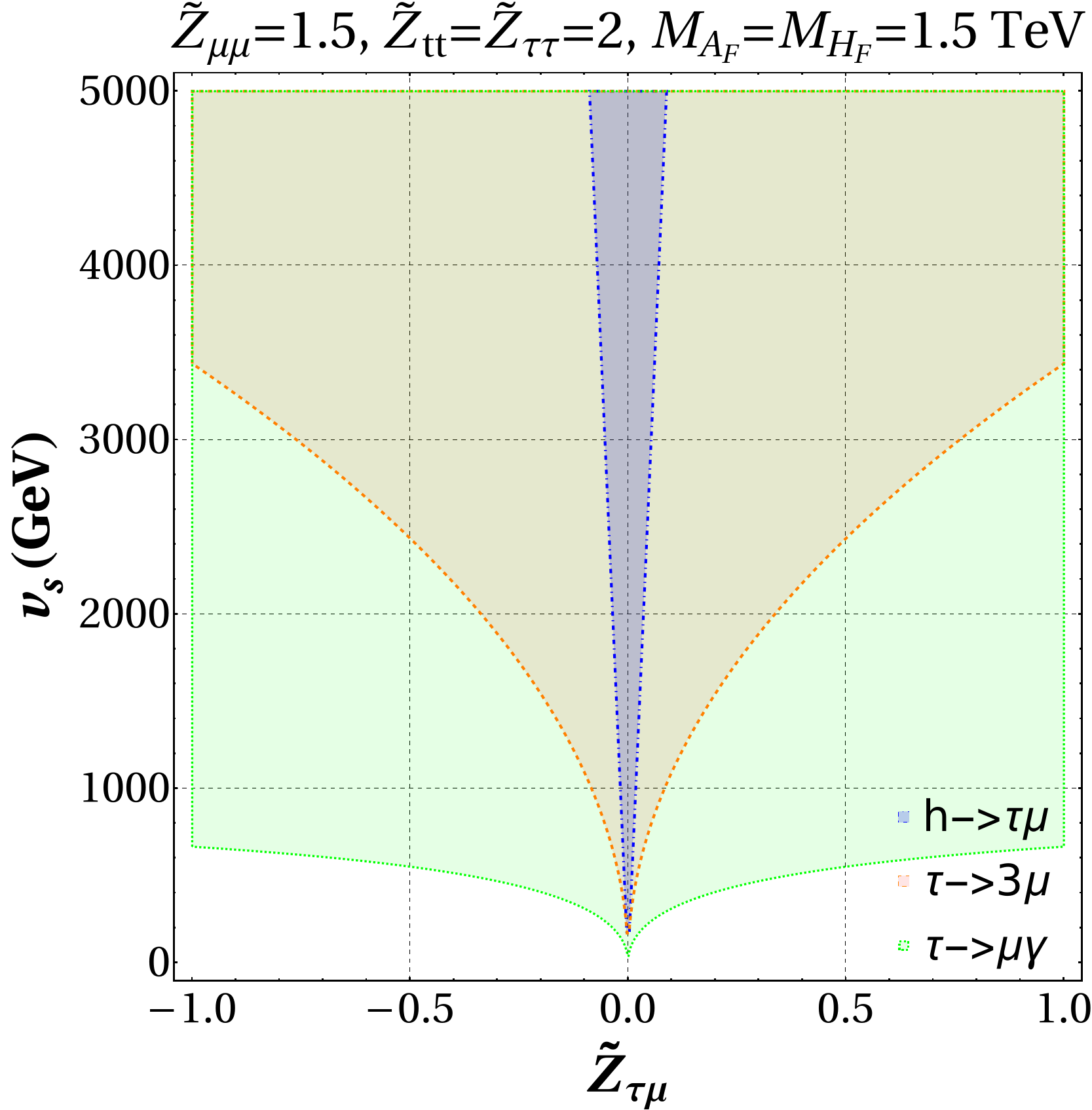}}
    \caption{(a) $\tilde{Z}_{e\mu}-v_s$ plane, and (b) $\tilde{Z}_{\tau\mu}-v_s$ plane. Different colored areas correspond to those allowed by upper limits on (a) $\mathcal{BR}(\mu\to e\gamma)$ and (b) $\mathcal{BR}(\tau\to \mu\gamma)$, $\mathcal{BR}(h\to \tau\mu)$, $\mathcal{BR}(\tau\to 3\mu)$. Additional parameters have been fixed, as explained in the main text. }\label{ParamSpace}
 \end{figure}

The parameter \( \tilde{Z}_{e\mu} \) is constrained by comparing its contributions to several LFV processes with their corresponding experimental upper limits. The relevant observables are the branching ratios of $\mu \to e \gamma$, $ \tau \to 3e$, and  $h \to e\mu$. The most stringent bound on \( \tilde{Z}_{e\mu} \) arises from the process \( \mu \to e \gamma \), due to the superior experimental sensitivity currently achieved for this decay mode. In particular, the direct search for the Higgs boson decay \( h \to e\mu \) yields a branching ratio constraint of order \( \mathcal{O}(10^{-5}) \), which is considerably weaker. This is primarily because the electron's small mass leads to a highly suppressed contribution to the decay width, as seen in Eq. \eqref{DeacyWid_Higgs-fermions}.
Thus, we show in Fig.~\ref{ParamSpace}(a) four regions in the $\tilde{Z}_{e\mu}-v_s$ plane compatible with the most stringent constraint: the upper limit on $\mathcal{BR}(\mu\to e\gamma)$ (where $\cos\alpha$ has been fixed at $0.995$)\footnote{This value ensures compatibility with the signal-strength modifiers $\mathcal{R}_X$ at the $1\sigma$ ($2\sigma$) level for $v_S \gtrsim  600\ \text{GeV}$ ($v_S \gtrsim 150\ \text{GeV}$).}. Each colored  allowed region corresponds to degenerate flavon masses $M_{A_F}=M_{H_F}=0.9,\ 1.5,\ 2.0$ and $2.5\ \text{TeV}$—consistent with earlier findings reported in Ref.~\cite{Arroyo-Urena:2022oft}—. In light of the current experimental upper limit on $\mathcal{BR}(\mu\to e\gamma)$, and considering an illustrative particular case with $v_s = 880\ \text{GeV}$\footnote{This value represents a very favorable scenario since it allows relatively large values for $\tilde{Z}_{e\mu}$.}, we obtain the following allowed intervals at:
\[
\begin{array}{rcl}
M_{A_F}=M_{H_F}=0.9\ \text{TeV}: & \quad & -0.0048 \lesssim \tilde{Z}_{e\mu} \lesssim 0.0048, \\[4pt]
M_{A_F}=M_{H_F}=1.5\ \text{TeV}: & \quad & -0.0130 \lesssim \tilde{Z}_{e\mu} \lesssim 0.0130, \\[4pt]
M_{A_F}=M_{H_F}=2.0\ \text{TeV}: & \quad & -0.0200 \lesssim \tilde{Z}_{e\mu} \lesssim 0.0200, \\[4pt]
M_{A_F}=M_{H_F}=2.5\ \text{TeV}: & \quad & -0.0260 \lesssim \tilde{Z}_{e\mu} \lesssim 0.0260. 
\end{array}
\]
Note that the couplings $g_{\phi\mu\mu}$ and $g_{\phi tt}$ in Eqs. \eqref{WilsonCoe1loop} and \eqref{WilsonCoe2loop} depend on $\tilde{Z}_{\mu\mu}$ and $\tilde{Z}_{tt}$, which were set to 1.5 and 2, respectively. In Table \ref{BRs_li-lj_a}, we present numerical estimates of the branching ratio $\mathcal{BR}(\mu\to e\gamma)$ within the FNSM\footnote{We present only the $\mu \to e \gamma$ channel, as its experimental bound is more stringent than those for $\tau \to e\gamma$ and $\tau \to \mu\gamma$.}, using benchmark parameter choices that directly influence the predicted rates. The values are compared to the current experimental upper limit, $\mathcal{BR}(\mu\to e\gamma)<4.2\times 10^{-13}$, demonstrating the viability of the selected parameter space under existing constraints. The computation includes both one- and two-loop contributions as given in Eqs.~\eqref{WilsonCoe1loop}-\eqref{WilsonCoe2loop}. As shown, the predicted branching ratios remain compatible with the experimental bound across the considered flavon mass range. The most stringent constraint arises from the case with lighter flavon masses $M_{H_F}=M_{A_F}=0.9$ TeV.

\begin{table}

\begin{centering}
\caption{Comparison of FNSM predictions for the branching ratios of the  $\mu \to e \gamma$, $\tau \to \mu \gamma$, $\tau \to e \gamma$ channels, at $\cos\alpha = 0.995$, and $v_s = 880$ GeV. The experimental upper bounds are given by $\mathcal{BR}(\mu \to e \gamma)<4.2\times 10^{-13}$, $\mathcal{BR}(\tau \to \mu \gamma)<4.2\times 10^{-8}$, $\mathcal{BR}(\tau \to e \gamma)<3.3\times 10^{-8}$.}\label{BRs_li-lj_a}
\begin{tabular}{|c|c|c|}
\hline 
$(\tilde{Z}_{e\mu}$, $\tilde{Z}_{\tau e}$, $\tilde{Z}_{\tau\mu}$) & $M_{H_{F}}=M_{A_{F}}$(GeV) & ($\mathcal{BR}(\mu\to e\gamma)$, $\mathcal{BR}(\tau\to e\gamma)$, $\mathcal{BR}(\tau\to \mu\gamma)$)\tabularnewline
\hline 
\hline 
(0.0048, 0.0001, 0.015) & 900 & ($3.08\times10^{-13}$, $1.5\times10^{-16}$, $3.39\times10^{-12}$)\tabularnewline
\hline 
(0.013, 0.0001, 0.015) & 1500 & ($2.97\times10^{-13}$, $2.2\times10^{-17}$, $5\times10^{-13}$) \tabularnewline
\hline 
(0.02, 0.0001, 0.015) & 2000 & ($2.75\times10^{-13}$, $8.6\times10^{-18}$, $1.9\times10^{-13}$) \tabularnewline
\hline 
(0.026, 0.0001, 0.015) & 2500 & ($2.91\times10^{-13}$, $4.6\times10^{-18}$, $1.04\times10^{-13}$) \tabularnewline
\hline 
\end{tabular}
\par\end{centering}
\end{table}

Constraints on $\tilde{Z}_{\tau\mu}$ are extracted from a combined analysis of three observables: $\tau \to 3\mu$, $\tau \to \mu \gamma$, and $h \to \tau \mu$.
Projected upper limits from the HL-LHC on $\mathcal{BR}(h\to\tau\mu) < 1 \times 10^{-4}$ yield the most stringent bounds, corresponding to
\begin{equation}
-0.015\lesssim \tilde{Z}_{\tau\mu} \lesssim 0.015, \, \text{at} \,v_s=880\,\text{GeV}.
\end{equation}

 It is important to highlight that quark-flavor constraints—specifically $B^0$–$\bar{B}^0$, $K^0$–$\bar{K}^0$, and $D^0$–$\bar{D}^0$ mixing—can impose stringent limits on certain model parameters. As illustrated in Refs.~\cite{Arroyo-Urena:2025zvg, Bauer:2016rxs, Abbas:2024dfh}, these constraints severely restrict the $M_{A_F}$–$v_s$ parameter plane, tightly bounding $M_{A_F}$ as a function of $v_s$. In our framework, we evade these restrictive bounds by suppressing the relevant matrix elements $\tilde{Z}_{db}$, $\tilde{Z}_{ds}$, and $\tilde{Z}_{uc}$, which respectively drive $B^0$–$\bar{B}^0$, $K^0$–$\bar{K}^0$, and $D^0$–$\bar{D}^0$ mixing at tree level. As demonstrated in Ref.~\cite{Arroyo-Urena:2025zvg} by one of the authors, the FNSM naturally accommodates all three mixing constraints simultaneously without tension. Crucially, the parameters governing these quark-flavor observables do not directly influence the phenomenological predictions in the presented work.
\\ 
 It is worth mentioning that in the FNSM, the mixing between the SM-like Higgs boson $h$ and the CP-even flavon $H_F$ is controlled by the angle $\alpha$, where the physical states are obtained from the gauge eigenstates by Eq.~\eqref{h-HF-mixing}. From Fig.~\ref{signal_strength}, current Higgs precision data constrain $\cos\alpha \approx 1$, implying very small mixing. In the exact limit $\cos\alpha = 1$ (and $\sin\alpha = 0$), the two CP-even scalars do not mix: $h$ coincides with the SM Higgs doublet component $\phi^0$, and $H_F$ corresponds to the real part of the flavon field $S_R$. In addition, there are several consequences for the Higgs phenomenology, namely, $i)$ Gauge and diagonal Yukawa couplings of $h$ become identical to the SM predictions, since $g_{hVV} \propto \cos\alpha$ and the diagonal fermion couplings follow $g_{hf\bar f} \simeq m_f/v$, $ii)$ LFV Higgs decays $h \to \ell_i\ell_j$ ($i\neq j$) vanishes, and $iii)$ Flavor-changing neutral currents (FCNC) would proceed only through the heavy flavon states $H_F$ and $A_F$.
 \\
 In conclusion, and for concreteness, we fix $c_\alpha = 0.995$ and express our results in the following sections in terms of the singlet complex VEV $v_s$, and the LFV couplings $\tilde{Z}_{e\mu}$ and $\tilde{Z}_{\tau\mu}$.


\section{Collider Analysis}\label{sec:SecIII}

In this section, we detail the collider analysis for the LFV Higgs decay channels $h \to \ell_i\ell_j$ in proton-proton collisions. The analysis employs a comprehensive Monte Carlo simulation framework to model both signal and background processes, and utilizing multivariate techniques --specifically Boosted Decision Trees (BDT) machine learning technique--, we achieve separation of the signal from the background.

Signal and background events were simulated with the following chain of stages: we first implement the full model using \texttt{FeynRules}~\cite{Alloul:2013bka} for compatibility with \texttt{MadGraph5}~\cite{Alwall:2014hca} considering the NNPDF2.3LO parton distribution functions~\cite{Ball:2012cx}. The generated events are then interfaced with \texttt{Pythia8}~\cite{Sjostrand:2014zea} for parton showering and hadronization, followed by detector simulation using \texttt{Delphes3}~\cite{deFavereau:2013fsa}. For this purpose, we employ the \texttt{delphes\_card\_HLLHC.tcl}~\cite{DelphesHLLHCCard} detector configuration for the HL-LHC.
\subsection{$h \to e\mu$ Channel}
\begin{itemize}
    \item \textbf{Signal:} This analysis focuses on the gluon-fusion production of the SM-like Higgs boson and its subsequent cLFV decay, $gg \to h \to e^\pm\mu^\mp$. Figure~\ref{cross_h-e_mu} presents the corresponding cross section as a function of the flavor-changing coupling $\tilde{Z}_{e\mu}$. By considering the benchmark point $M_{H_F}=M_{A_F}=0.9$ TeV and $v_s=1$ TeV, the green band highlights the parameter space allowed by the most stringent experimental constraint, derived from the current upper limit on $\mathcal{BR}(\mu \to e\gamma)$. This constraint imposes $\tilde{Z}_{e\mu} \lesssim 0.005$, which directly translates to an upper bound on the signal cross section of $\sigma(gg \to h \to e\mu) \lesssim 0.00076$ pb. Given this maximal cross section, the High-Luminosity LHC (HL-LHC) at its ultimate integrated luminosity of $3\ \text{ab}^{-1}$ could yield up to 2280 signal events. This projected event count represents a promising opportunity to probe the cLFV Higgs decay channel $h \to e\mu$, potentially offering sensitivity to new physics in the charged lepton sector.  
    
    \begin{figure}[!htb]
    \centering    \includegraphics[scale=0.4]{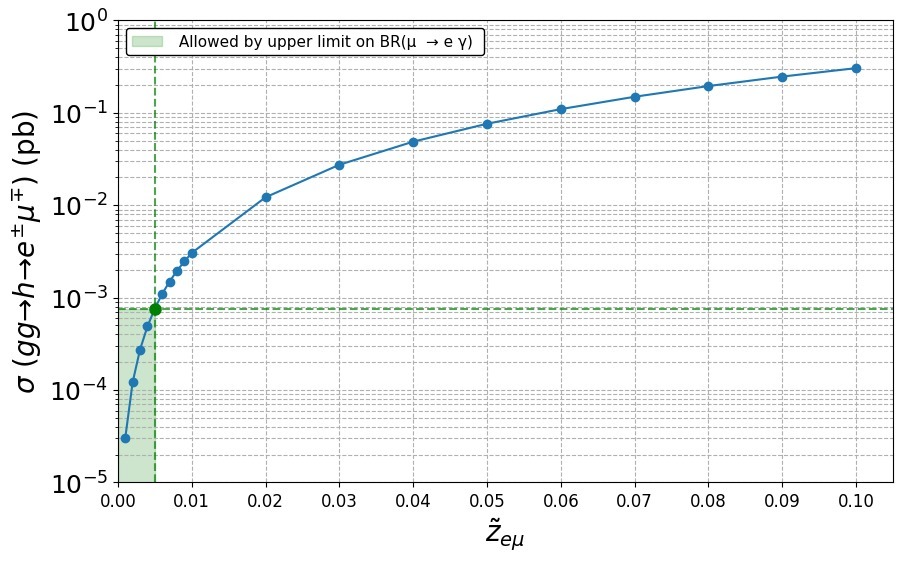}
    \caption{Production cross section of the signal as a function of $\tilde{Z}_{e\mu}$. 
    The green region represents that allowed by the upper bound on $\mathcal{BR}(\mu\to e\gamma)$.}
    \label{cross_h-e_mu}
\end{figure}

    \item \textbf{Background:} While SM background processes can yield final states with different-flavor leptons, they originate from mechanisms fundamentally distinct from those involving an explicit lepton-flavor-violating vertex within a single decay. Such backgrounds typically involve undetected neutrinos, manifesting experimentally as missing transverse energy. Consequently, their kinematic distributions—particularly in the region near a resonance in the 
$e\mu$ invariant mass—deviate significantly from the sharp, two-body decay signature expected from  \(h \to e\mu\). The dominant background processes contributing to this channel include:
    \begin{itemize}
    \item \textit{Drell-Yan Production:} $pp\to Z/\gamma^* \to \tau^+\tau^-$ with leptonic tau decays ($\tau \to e\nu_e\nu_\tau$, $\tau \to \mu\nu_\mu\nu_\tau$),
    \item \textit{Top Quark Pair Production:} $pp\to t\bar{t}$ decays yielding opposite-sign $e\mu$ pairs through semileptonic decays,
    \item \textit{Diboson Production:} 
    \begin{itemize}
        \item $W^+W^-$ with $W^+ \to e^+\nu_e$, $W^- \to \mu^-\bar{\nu}_\mu$,
        \item $WZ$ with $W \to e\nu$, $Z \to \mu\mu$ (one muon missed),
        \item $ZZ$ with both $Z$ bosons decaying leptonically.
    \end{itemize}
\end{itemize}
\end{itemize}
The production cross sections of the background processes are presented in Table \ref{tab:bkg-xsecs_h-e_mu}.
\begin{table}[ht]
  \centering
     \begin{tabular}{l c}
    \hline
    \textbf{Background Process} & \textbf{Cross Section [pb]} \\
    \hline
    Drell--Yan            & $26$ \\
    Top Quark Production  & 127.7 \\
    \hline
    \multicolumn{2}{c}{\emph{Diboson Production}} \\
    \hline
    $W^{+}W^{-}$          & 0.8378 \\
    $WZ$                  & 0.1364 \\
    $ZZ$                  & 0.09846 \\
    \hline
  \end{tabular}
   \caption{Cross sections for each background process for $h \to e\mu$.}\label{tab:bkg-xsecs_h-e_mu}
\end{table}

\subsection{$h \to \tau^\pm\mu^\mp$ ($\tau^\pm \to \pi^\pm \nu_\tau$) channel}

\begin{itemize}
    \item \textbf{Signal:} In this channel, we search for a final state including a charged lepton $\mu=\mu^-,\,\mu^+$, a charged pion $\pi^{\pm}$ and missing transverse energy due to undetected neutrinos: $gg \to h \to \tau^\pm\mu^\mp$ with $\tau^\pm \to \pi^\pm \nu_\tau$.
 The signal production cross section as a function of $\tilde{Z}_{\tau\mu}$ is presented in Fig.~\ref{cross_h-tau_mu}. 
  
    \begin{figure}[!htb]
    \centering    \includegraphics[scale=0.4]{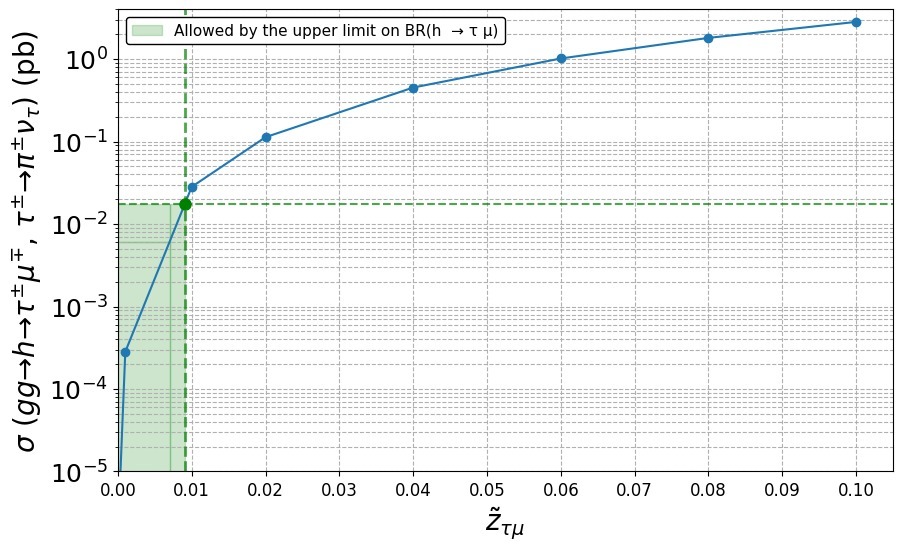} 
    \caption{
    Production cross section of the signal as a function of $\tilde{Z}_{\tau\mu}$.  
    The green area indicates the allowed region by the direct upper limit on $\mathcal{BR}(h\to \tau\mu)$.}
    \label{cross_h-tau_mu}
\end{figure}
As with the prior channel, the green region delineates the parameter space compatible with the most stringent experimental constraint, which stems from the current upper limit on $\mathcal{BR}(h \to \tau\mu)$. This bound imposes $\tilde{Z}_{\tau\mu} \lesssim 0.007$. Within the model, this coupling limit translates to a maximal cross section of $\sigma(gg \to h \to \tau\mu) \lesssim 0.0061\ \text{pb}$, which corresponds to an indirect upper bound of $\mathcal{BR}(h \to \tau\mu) \lesssim 1 \times 10^{-5}$. This theoretical prediction is consistent with the direct experimental limit and is, in fact, more restrictive by approximately two orders of magnitude.
\item \textbf{Background:}

    \begin{itemize}
    \item \textit{Drell-Yan Production:} $pp\to Z/\gamma^* \to \tau^+\tau^-$, which leads to the decay channels:
    \begin{itemize}
    \item \textit{Leptonic tau decays:} $\tau \to \ell\nu_\ell\nu_\tau$,
    \item \textit{Semileptonic tau decays:} $\tau \to \ell\nu_\ell\nu_\tau$ and $\tau^\pm \to \pi^\pm \nu_\tau$,
    \item \textit{Hadronic tau decays:} $\tau^\pm \to \pi^\pm \nu_\tau$,
    \end{itemize}
    \item \textit{Top Quark Pair Production:} $t\bar{t}$ decays yielding muons and jets,
    \item \textit{$WZ$:} $W \to jj$ and $Z \to \ell^{+}\ell^{-}$,
    \item \textit{Leptonic $WW$ decays:} $W \to \ell\nu_\ell$,
    \item \textit{Semileptonic $WW$ decays:} One $W$ decays leptonically $W \to \ell\nu_\ell$, and the other $W \to jj$.
\end{itemize}
\end{itemize}
The production cross sections of the background processes are presented in Table \ref{tab:bkg-xsecs_h-tau_mu}.
\begin{table}[ht]
  \centering
  \begin{tabular}{l c}
    \hline
    \textbf{Background Process} & \textbf{Cross Section [pb]} \\
    \hline
    Top Quark Production  & 164.8 \\
    \hline
    \multicolumn{2}{c}{\emph{Diboson Production}} \\
    \hline
    $W^{+}W^{-}$ (leptonic decay)      & 3.4 \\
    $W^{+}W^{-}$ (semileptonic decay)  & 30.9 \\
    $WZ$                  & 1.2 \\
    $ZZ$                  & 0.1 \\
    \hline
    \multicolumn{2}{c}{\emph{Drell-Yan}} \\
    \hline
    Leptonic Decays         & $115.6$ \\
    Semileptonic Decays     & $35.5$ \\
    Hadronic  Decays        & $10.9$ \\
    \hline
  \end{tabular}
   \caption{Cross sections for each background process for $h \to \tau \mu$.}   \label{tab:bkg-xsecs_h-tau_mu}

\end{table}


\subsection{Multivariate Analysis}
Once the kinematic analysis is completed, we notice that most of the observables used to distinguish the signal from the background have relatively weak discriminating relevance. Thus, we opted to implement selection by determining Multivariate Analysis (MVA) discriminators. These discriminators combine the observables into a single, more powerful classifier. For the MVA training, we use a Boosted Decision Tree (BDT) method \cite{Coadou:2022nsh}, implemented via the XGBoost library \cite{Chen:2016:XST:2939672.2939785}, using an advanced gradient boosting technique. We trained the BDT classifiers with kinematic observables from the final state particles, including the invariant mass, transverse momentum ($p_T$), pseudorapidity ($\eta$), azimuthal angle ($\phi$), the missing transverse energy, angular separation ($\Delta R$), etc. The BDT training is performed using MC-simulated samples.
The signal and background samples are scaled to the expected number of candidates, computed from the integrated luminosity and cross sections. The optimization of the BDT selection is performed separately for each channel to enhance the figure of merit, specifically the signal significance, defined as $Z=S/\sqrt{S + B+(x\cdot B)^2}$. Here, $S$ and $B$ denote the number of signal and background candidates, respectively. The factor $x$ stands for a systematic uncertainty. The main experimental systematic uncertainties considered include 
the integrated luminosity $(1\%)$, lepton identification efficiencies 
$(1.5\%)$, trigger efficiencies $(1\%)$, and momentum scale and resolution 
$(0.7\%)$. Theoretical uncertainties account for QCD scale variations $(2\%)$, 
parton distribution functions $(1.5\%)$, and background modeling $(2\%)$. 
The multivariate analysis classifier adds an additional $2\%$ uncertainty. 
These contributions are combined in quadrature, yielding a total systematic 
uncertainty of $4.36\%$, consistent with HL-LHC projections for analyses 
involving well-reconstructed leptonic final states~\cite{CMS:2018far}. 

\begin{figure}[!htb]
    \centering   
     \subfigure[]{\includegraphics[scale=0.315]{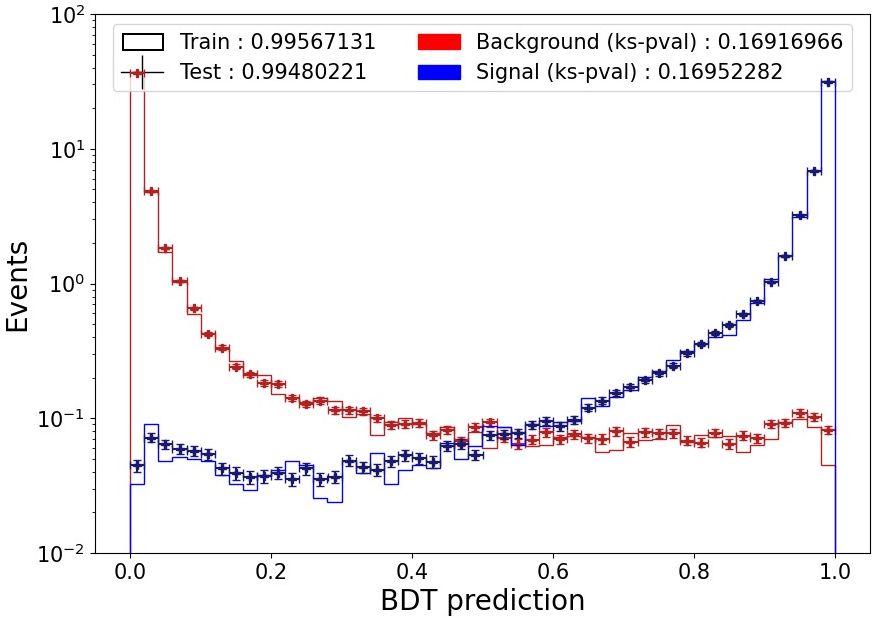}}
          \subfigure[]{\includegraphics[scale=0.315]{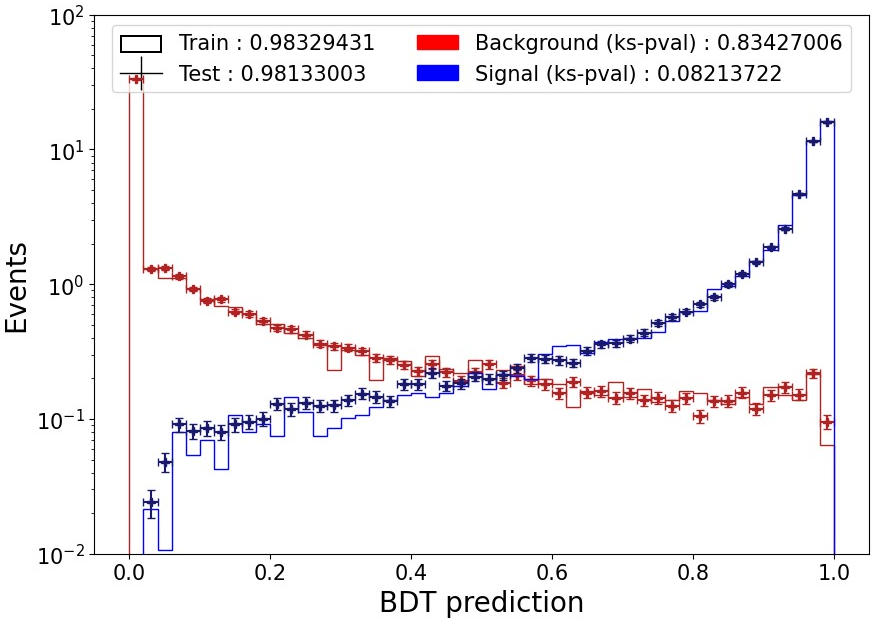}}
    \caption{Plot of the discriminant for signal and
background data: (a) $h\to e\mu$ channel and (b) $h\to \tau\mu$ channel.}
    \label{KSplot}
\end{figure}

 The BDT model was optimized with the Optuna framework \cite{akiba2019optuna} and its performance was assessed using the Kolmogorov-Smirnov (KS) test. The discriminant distributions for signal and background are shown in Fig. \ref{KSplot}. The computed KS statistics, all within the acceptable [0.05, 1] range, are 0.17 (both signal and background) for the $h\to e\mu$ channel and 0.83 (background) and 0.08 (signal) for the $h\to \tau\mu$ channel. This demonstrates that the training successfully mitigated overtraining, resulting in a model with strong generalizability.

\subsection{Results}

\subsection*{$h\to e\mu$ channel}
\begin{figure}[!htb]
    \centering   
    \subfigure[]{\includegraphics[scale=0.235]{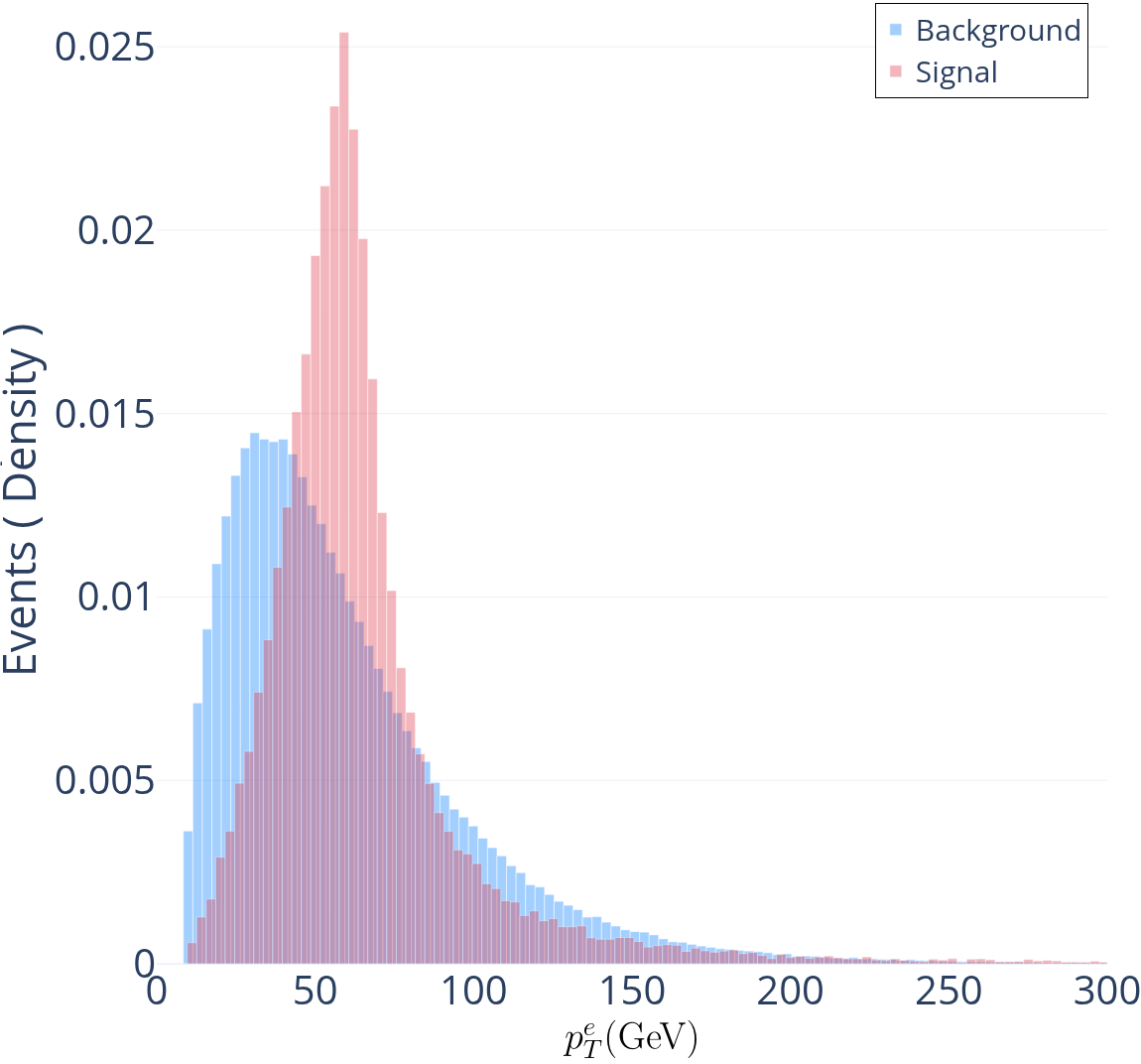}}
    \subfigure[]{\includegraphics[scale=0.235]{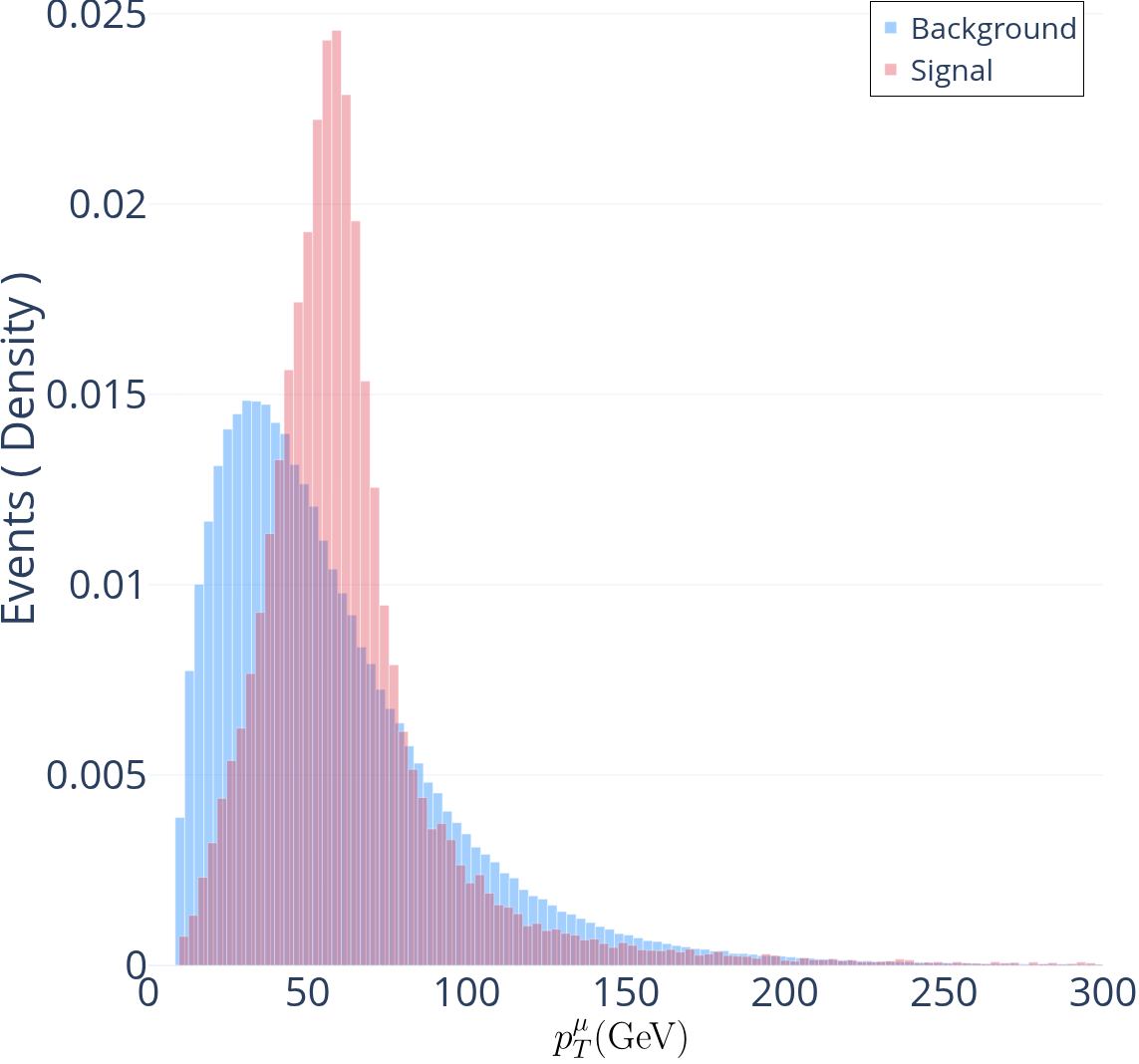}}
    \caption{Kinematical distributions of the most discriminant observables to separate the signal from the background. (a) Transverse momentum of the electron $p_T^e$, and (b) transverse momentum of the muon $p_T^{\mu}$.}
    \label{pTs_e-mu}
\end{figure}

Despite experimental challenges, this channel is anticipated to yield a clear signature given sufficient integrated luminosity. Key discriminating observables are the electron and muon transverse momentum ($p_T^e$ and $p_T^\mu$), which are predicted to peak at approximately $m_h/2$ GeV in a significant fraction of events, as shown in Fig. \ref{pTs_e-mu}. The successful measurement of the decay of $h \to \mu^-\mu^+$ provides encouragement for the search of $h \to e\mu$ as the coupling of $g_{he\mu} \sim 1.7 \times 10^{-5}$ is only one order of magnitude smaller than $g_{h\mu\mu}$, a disadvantage that could be compensated  by the high luminosity of the HL-LHC.

\begin{figure}[!htb]
    \centering    
    \subfigure[]{\includegraphics[scale=0.2]{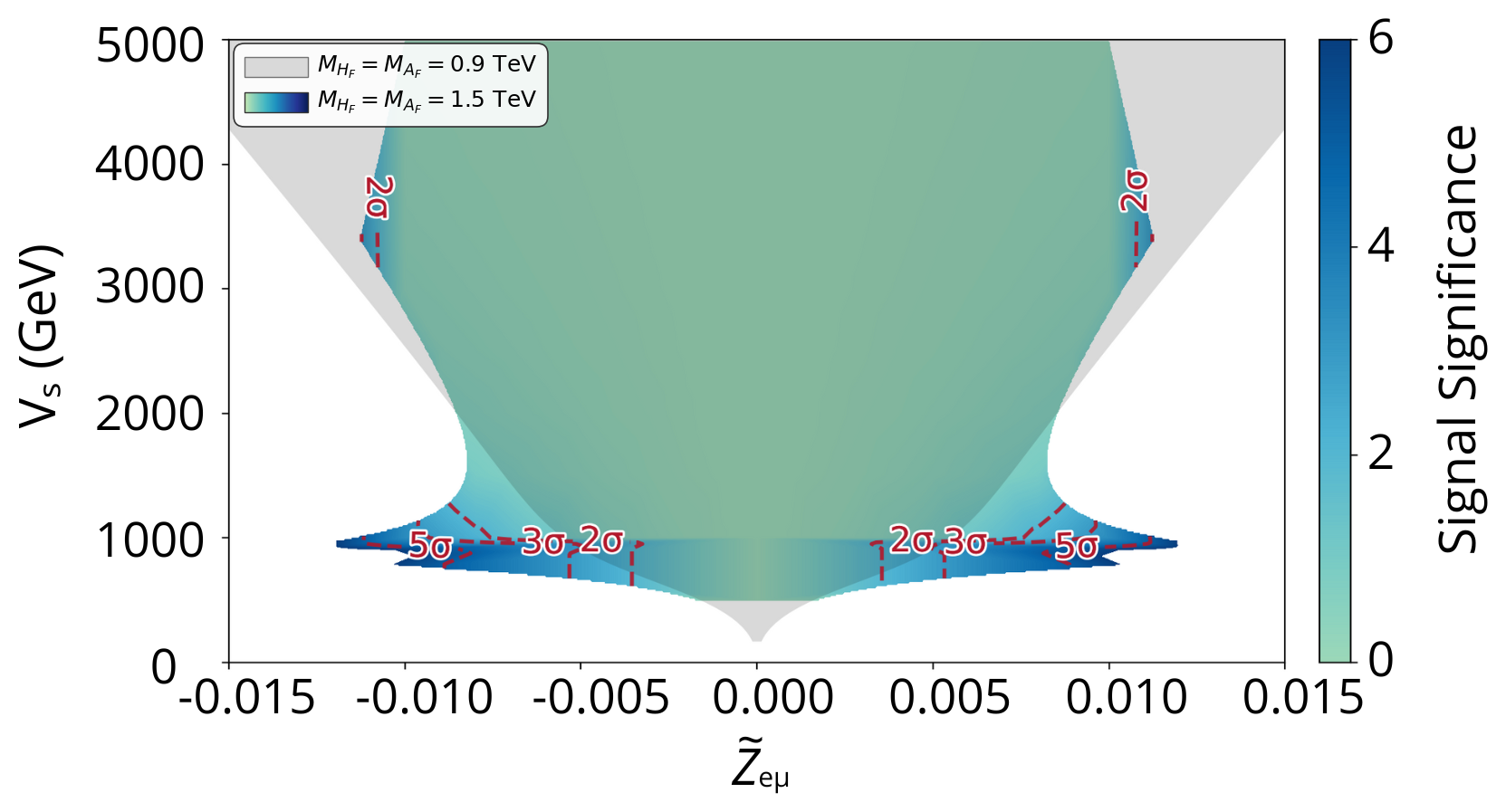}}
     \subfigure[]{\includegraphics[scale=0.2]{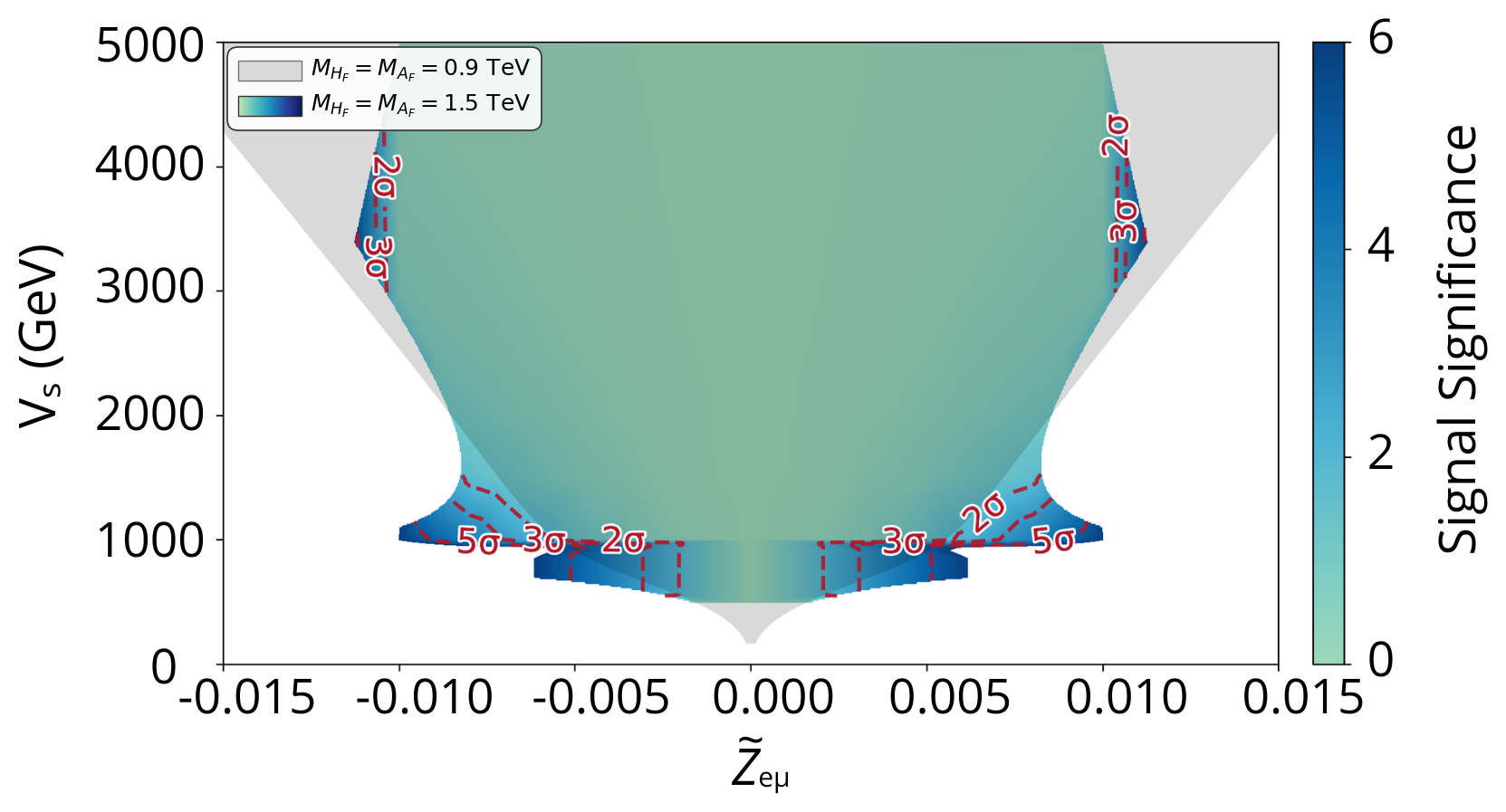}}
    \caption{Signal significance as a function of the singlet complex vev $v_s$ and lepton flavor-violating parameter $\tilde{Z}_{e\mu}$ for integrated luminosities of 1000 and 3000 fb$^{-1}$, and a cut on XGB of 0.995.}
    \label{signvslumhemu}
\end{figure}

Figure~\ref{signvslumhemu} presents our central phenomenological results: the projected signal significance for the cLFV decay $h \to e\mu$ as a function of two fundamental Froggatt-Nielsen model parameters, namely, the complex singlet vacuum expectation value $v_{s}$ and the flavor-changing coupling $\tilde{Z}_{e \mu}$.

 The significance is computed for integrated luminosities of $1000\ \text{fb}^{-1}$ and $3000\ \text{fb}^{-1}$, applying a BDT cut of 0.95 to maximize the signal-to-background ratio. As anticipated, the significance increases with larger values of $\tilde{Z}_{e \mu}$, a direct consequence of the partial width scaling $\Gamma (h \to e\mu) \propto \tilde{Z}_{e \mu}^{2}$. Conversely, it diminishes with increasing $v_{s}$, since all LFV couplings are universally suppressed by $v_{s}^{-1}$.  We delineate two primary regions of interest, corresponding to benchmark heavy scalar masses $M_{H_F}=M_{A_F}=0.9\ \text{TeV}$ and $1.5\ \text{TeV}$, represented by a gray contour and a green-blue color palette, respectively\footnote{If one considers the two additional benchmark scenarios for the $h \to e\mu$ channel—specifically, $M_{H_F} = M_{A_F} = 2\ \text{TeV}$ and $2.5\ \text{TeV}$—even higher signal significances would be expected. This is due to the more permissible parameter space available for larger heavy scalar masses, as clearly illustrated in the allowed region plot (Fig. \ref{ParamSpace}(a)).}. For the $1.5\ \text{TeV}$ case (color palette), our analysis reveals that, within the parameter intervals $(-0.012,-0.008) \lesssim \tilde{Z}_{e \mu} \lesssim (0.008, 0.012)$ and $v_s \approx 900\ \text{GeV}$, an integrated luminosity of $\mathcal{L}_{\rm int} = 1000\ \text{fb}^{-1}$ yields a projected significance of approximately $5\sigma$. This result suggests the feasibility of obtaining the first experimental evidence for this cLFV Higgs decay channel. Scaling the luminosity to $\mathcal{L}_{\rm int} = 3000\ \text{fb}^{-1}$, tightens the coupling intervals to $(-0.011,-0.007) \lesssim \tilde{Z}_{e \mu} \lesssim (0.007, 0.011)$, reflecting an enhanced precision in the prospective measurement. The gray contour, corresponding to the lighter $0.9\ \text{TeV}$ case, generally projects lower significances. For instance, it reaches the $3\sigma$ threshold for $|\tilde{Z}{e \mu}| \approx 0.005$ at $\mathcal{L}{\rm int} = 1000\ \text{fb}^{-1}$. This region is therefore less favored for couplings of order $\tilde{Z}_{e \mu}\sim\mathcal{O}(0.001)$.
  

\subsection*{$h \to \tau\mu$ channel}

\begin{figure}[!htb]
    \centering   
    \subfigure[]{\includegraphics[scale=0.235]{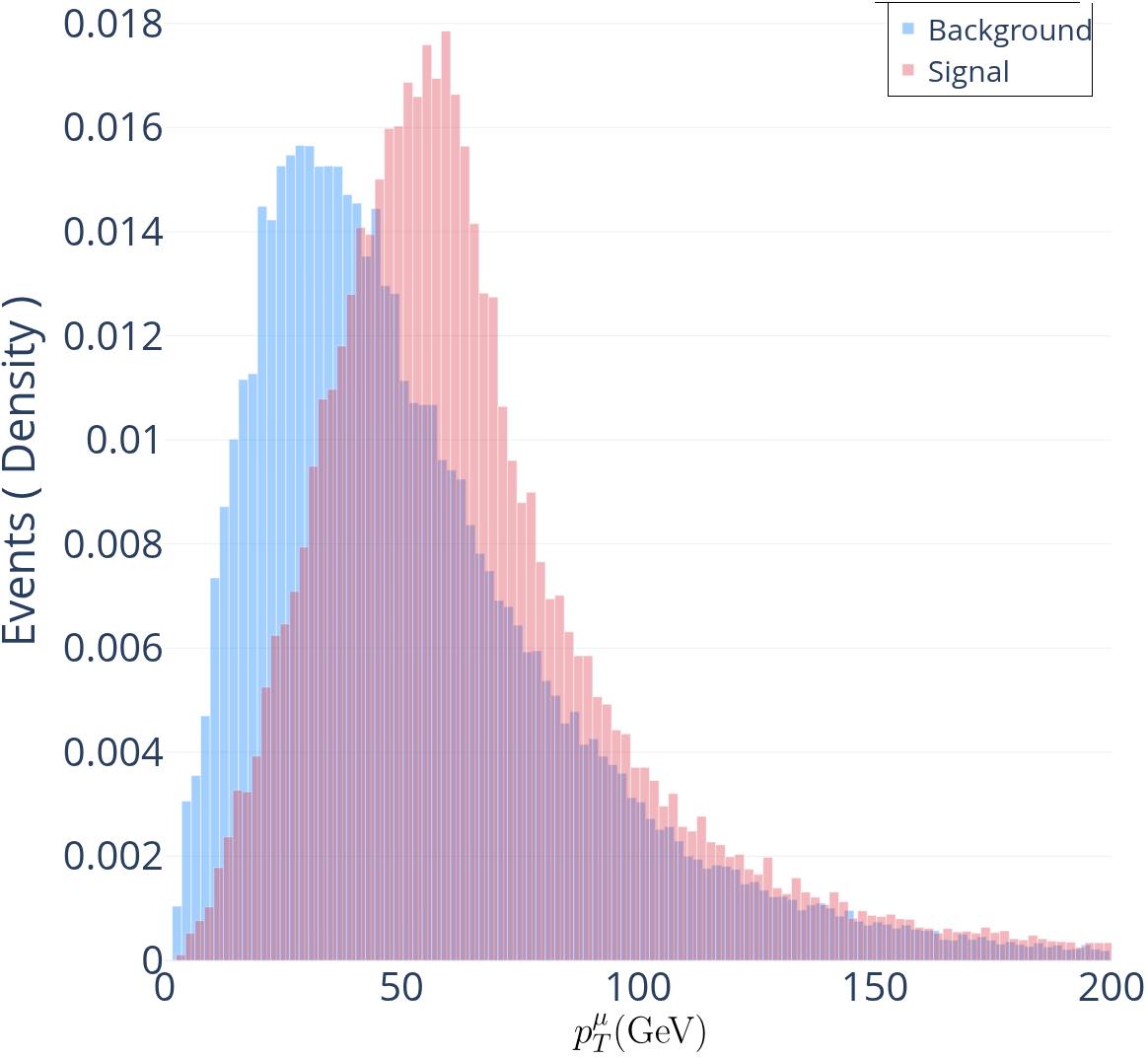}}
    \subfigure[]{\includegraphics[scale=0.235]{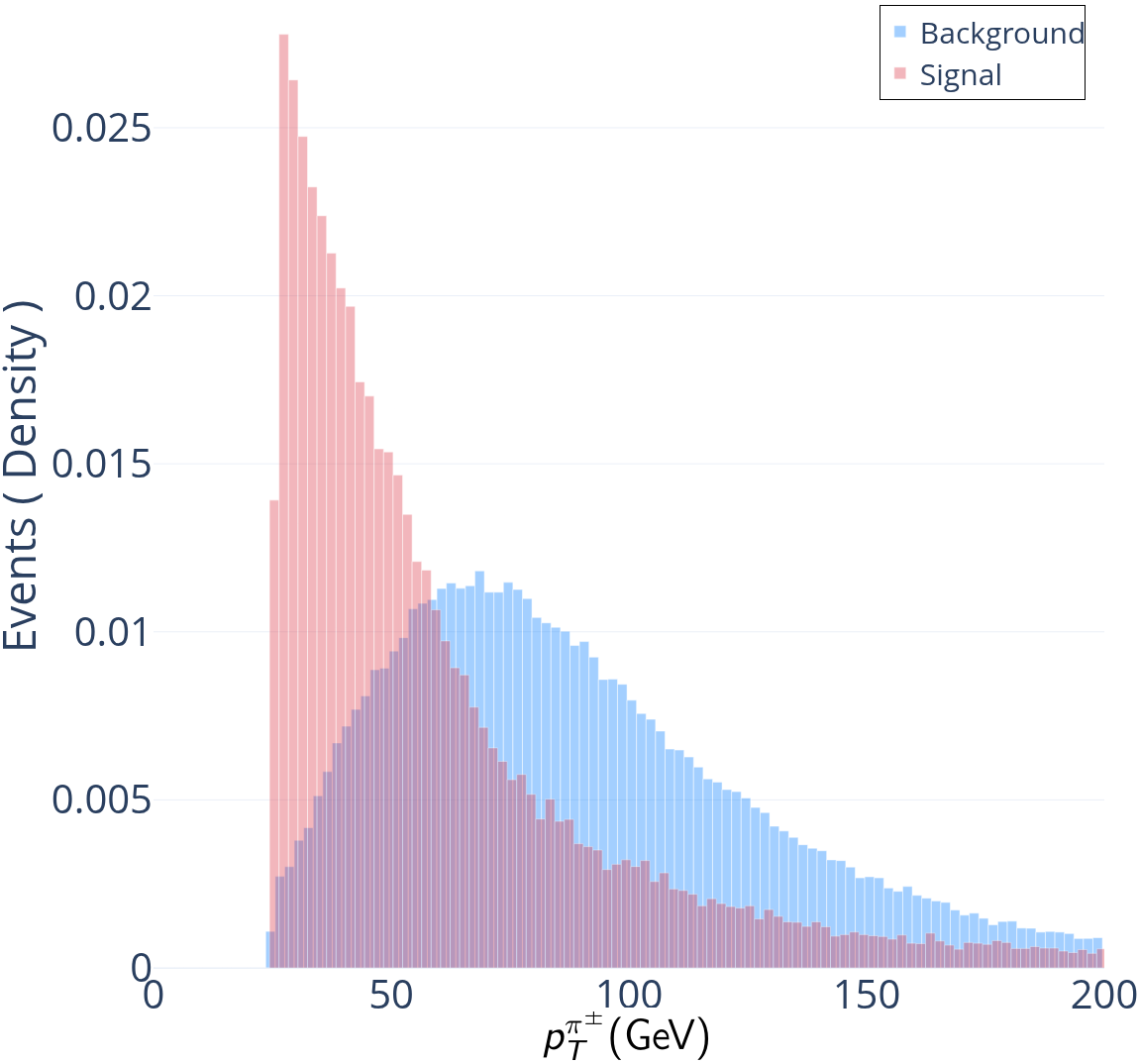}}
    \caption{Kinematical distributions of the most discriminant observables to separate the signal from the background. (a) Transverse momentum of the muon $p_T^\mu$, and (b) transverse momentum of the charged pion $p_T^{\pi^\pm}$.}
    \label{pTshtaumu}
\end{figure}

\begin{figure}[!htb]
    \centering   
    \subfigure[]{\includegraphics[scale=0.195]{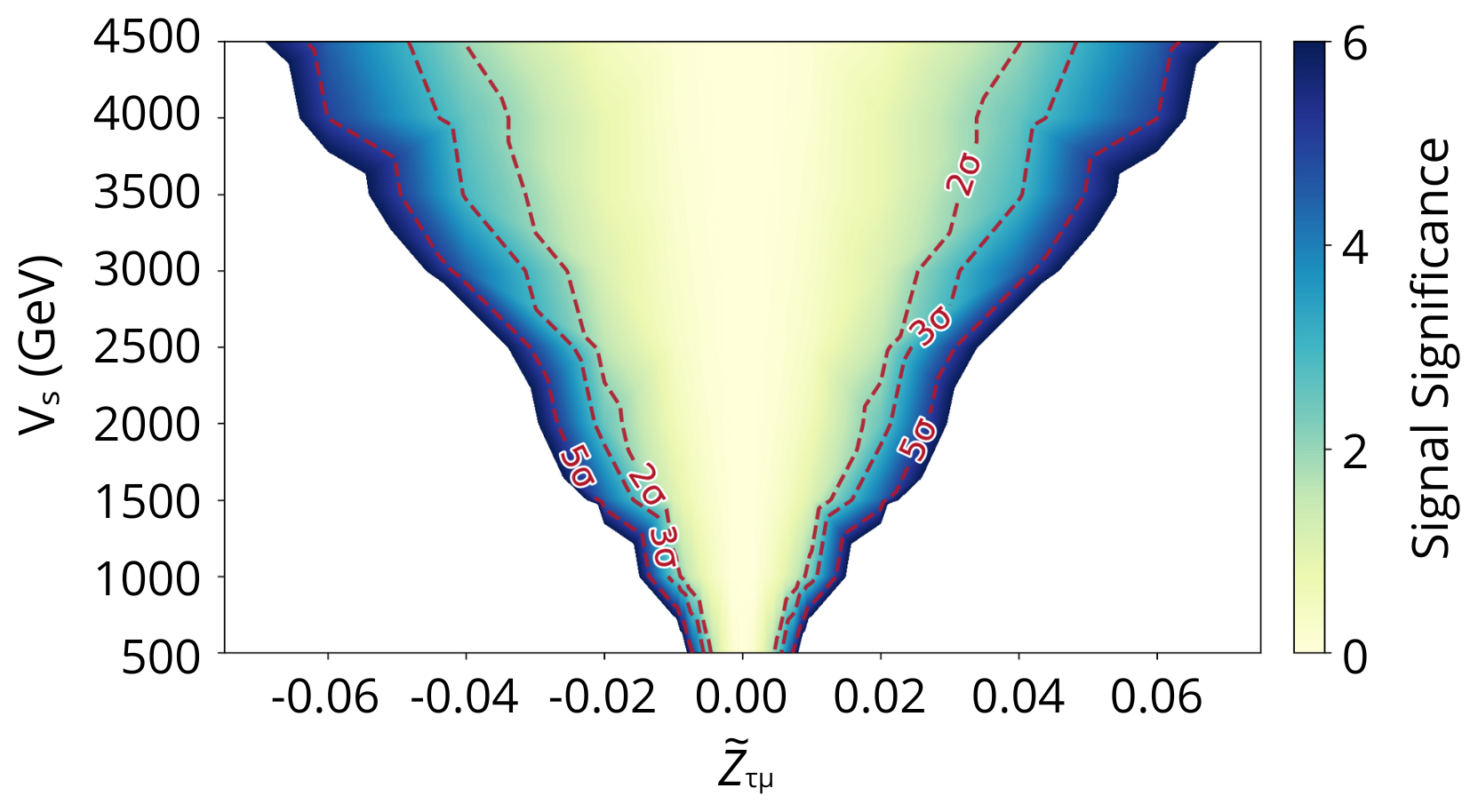}}
    \subfigure[]{\includegraphics[scale=0.195]{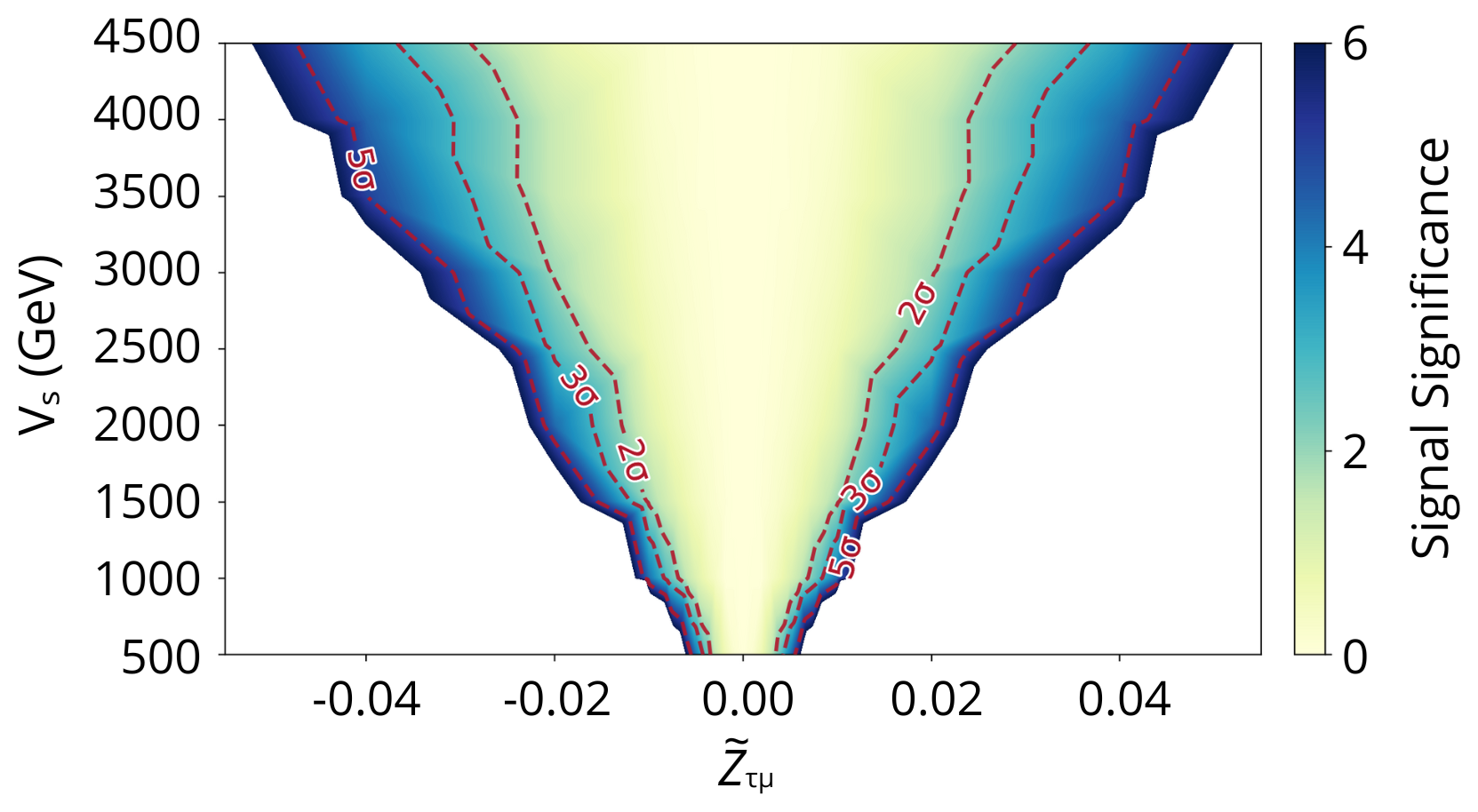}}
    \caption{Signal significance as a function of the singlet complex vev $v_s$ and lepton flavor-violating parameter $\tilde{Z}_{\tau\mu}$ for integrated luminosities of 1000 and 3000 fb$^{-1}$, and a cut on XGB of 0.95.}
    \label{signvslumhtaumu}
\end{figure}

    Following the same approach as the previous channel, we present in Fig.~\ref{pTshtaumu} the most important observables to separate the signal from the background processes: the muon and pion transverse momentum ($p_T^\mu$ and $p_T^{\pi^\pm}$). Meanwhile, Fig.~\ref{signvslumhtaumu} shows the projected significance for the cLFV decay channel $h\to\tau\mu$ as a function of the VEV $v_s$ and the flavor-changing coupling $\tilde{Z}_{\tau\mu}$. The contours correspond to a projected integrated luminosity of (a) 1000 fb$^{-1}$ and (b) 3000 fb$^{-1}$. Our findings reveal that the HL-LHC could also probe this channel with a significance at $5\sigma$ level for $-0.01\lesssim Z_{\tau\mu}\lesssim 0.01$ ($-0.006\lesssim Z_{\tau\mu}\lesssim 0.006$) and $v_s=500$ GeV, once the integrated luminosity exceeds a value of 1000 fb$^{-1}$ (3000 fb$^{-1}$). For this channel we also applied a BDT cut of 0.95.

\subsection*{$h \to \tau e$ channel}

Within the Froggatt-Nielsen model, the $h\tau e$ interaction is directly governed by the $\tilde{Z}_{\tau e}$ coupling. Our analysis of the parameter space reveals that the individual observables—$\tau\to e\gamma$, $\tau\to3e$, and $h\to \tau e$—are not, on their own, strict enough to impose stringent bounds on $\tilde{Z}_{\tau e}$, which can be as large as $\mathcal{O}(1)$. However, a value of this magnitude could be unnatural, given that $\tilde{Z}_{\tau \mu}$ and $\tilde{Z}_{e \mu}$ are in the range $0-\mathcal{O}(10^{-2})$. Motivated by the hierarchy in the quark and lepton sectors, where the CKM and PMNS matrix elements for mixing between the first and third families are smaller than those between the first and second or the second and third, we instead postulate the hierarchy $\tilde{Z}_{\tau e} < \tilde{Z}_{e\mu} < \tilde{Z}_{\tau \mu}$. Using this naturalness argument as a benchmark, we estimate $\tilde{Z}_{\tau e}$ should be less than $\mathcal{O}(10^{-4})$. For a value of $\tilde{Z}_{\tau e} = 0.0005$, we find a cross section of $\sigma(gg\to h\to\tau e, \tau\to \mu\nu_\mu \nu_\tau)=9.27 \times 10^{-10}$ pb. This is very small compared to the $h \to e\mu$ and $h \to \tau\mu$ channels, which have cross sections of order $\lesssim10^{-3}$ pb. Consequently, the predicted number of events for the $\tau e$ channel would be significantly smaller and undetectable, even at the integrated luminosity of $3000$ fb$^{-1}$.
\section{Conclusions}\label{sec:SecIV}

We have conducted a comprehensive analysis of the discovery prospects for charged lepton flavor-violating Higgs boson decays, \( h \to \ell_i \ell_j \) (\(\ell_i \neq \ell_j\)), within the Froggatt-Nielsen Singlet Model at the High-Luminosity LHC.

By systematically scanning the model's phenomenologically viable parameter space--constrained by Higgs precision data and stringent low-energy limits on processes like \(\ell_i \to \ell_j\gamma\)--we identified benchmark scenarios potentially accessible at the HL-LHC. Our collider analysis, based on Monte Carlo simulations and advanced multivariate techniques (Boosted Decision Trees), projects the experimental sensitivity for the most promising channels.

Our main findings are as follows:
\begin{itemize}
    \item The \( h \to e\mu \) channel could reach the \(5\sigma\) discovery threshold with an integrated luminosity of \( \mathcal{L}_{\text{int}} \gtrsim 1000~\text{fb}^{-1} \), for flavor-violating couplings in the range \( |\tilde{Z}_{e\mu}| \sim \mathcal{O}(10^{-2}) \) and a singlet VEV \( v_s \sim 900~\text{GeV} \).
    \item Similarly, the \( h \to \tau\mu \) channel (with \( \tau^\pm \to \pi^\pm \nu_\tau \)) is also projected to be discoverable at the \(5\sigma\) level for \( \mathcal{L}_{\text{int}} \gtrsim 1000~\text{fb}^{-1} \), given a coupling \( |\tilde{Z}_{\tau \mu}| \sim \mathcal{O}(10^{-3}) \)  for $v_s=500$ GeV.
    \item In contrast, the \( h \to \tau e \) decay is predicted to remain unobservable even at the ultimate HL-LHC luminosity of \( 3000~\text{fb}^{-1} \). This is a consequence of the natural hierarchy \( \tilde{Z}_{\tau e} < \tilde{Z}_{e\mu} < \tilde{Z}_{\tau \mu} \), which leads to a negligibly small production cross section.
\end{itemize}

These results underscore the unique and complementary role of the HL-LHC as a probe of new physics. While low-energy experiments provide exceptionally tight constraints on the parameter space of BSM models, the direct search for rare Higgs decays offers a powerful, high-energy window into flavor dynamics. The potential observation of \( h \to e\mu \) or \( h \to \tau\mu \) would constitute a smoking-gun signature of physics BSM, providing direct evidence for a non-trivial flavor structure in the Higgs sector. Conversely, the null results projected for \( h \to \tau e \) within our framework present a specific, testable prediction of the Froggatt-Nielsen mechanism's flavor hierarchy at the energy frontier.

\section*{Acknowledgments}
{The work of M. A. Arroyo-Ure\~na is supported by “Estancias
posdoctorales por México (SECIHTI)”, “Sistema Nacional de Investigadoras e Investigadores” (SNII), and the SECIHTI project No. CBF-2025-G-1187. E. A. Herrera-Chacón would like to thank Frank Krauss for useful initial discussions. S. Rosado-Navarro thanks VIEP-BUAP for the support. }

\bibliographystyle{JHEP}
\bibliography{references}

@article{Coadou:2022nsh,
    author = "Coadou, Yann",
    title = "{Boosted decision trees}",
    eprint = "2206.09645",
    archivePrefix = "arXiv",
    primaryClass = "physics.data-an",
    doi = "10.1142/9789811234033_0002",
    month = "3",
    year = "2022"
}

@article{Belle:2021ysv,
    author = "Abdesselam, A. and others",
    collaboration = "Belle",
    title = "{Search for lepton-flavor-violating tau-lepton decays to $\ell\gamma$ at Belle}",
    eprint = "2103.12994",
    archivePrefix = "arXiv",
    primaryClass = "hep-ex",
    doi = "10.1007/JHEP10(2021)019",
    journal = "JHEP",
    volume = "10",
    pages = "19",
    year = "2021"
}

@article{Cepeda:2019klc,
    author = "Cepeda, M. and others",
    editor = "Dainese, Andrea and Mangano, Michelangelo and Meyer, Andreas B. and Nisati, Aleandro and Salam, Gavin and Vesterinen, Mika Anton",
    title = "{Report from Working Group 2: Higgs Physics at the HL-LHC and HE-LHC}",
    eprint = "1902.00134",
    archivePrefix = "arXiv",
    primaryClass = "hep-ph",
    reportNumber = "CERN-LPCC-2018-04",
    doi = "10.23731/CYRM-2019-007.221",
    journal = "CERN Yellow Rep. Monogr.",
    volume = "7",
    pages = "221--584",
    year = "2019"
}

@article{Blankenburg:2012ex,
    author = "Blankenburg, Gianluca and Ellis, John and Isidori, Gino",
    title = "{Flavour-Changing Decays of a 125 GeV Higgs-like Particle}",
    eprint = "1202.5704",
    archivePrefix = "arXiv",
    primaryClass = "hep-ph",
    reportNumber = "CERN-PH-TH-2012-049, KCL-PH-TH-2012-09, LCTS-2012-05",
    doi = "10.1016/j.physletb.2012.05.007",
    journal = "Phys. Lett. B",
    volume = "712",
    pages = "386--390",
    year = "2012"
}

@techreport{CMS:2018far,
    author = "{CMS Collaboration}",
    title = "{CMS Phase II Upgrade Scope Document}",
    institution = "CERN",
    number = "CERN-LHCC-2018-019",
    year = "2018",
    url = "https://cds.cern.ch/record/2643065"
}

@article{ParticleDataGroup:2024cfk,
    author = "Navas, S. and others",
    collaboration = "Particle Data Group",
    title = "{Review of particle physics}",
    doi = "10.1103/PhysRevD.110.030001",
    journal = "Phys. Rev. D",
    volume = "110",
    number = "3",
    pages = "030001",
    year = "2024"
}

@inproceedings{akiba2019optuna,
    title = "{Optuna: A Next-generation Hyperparameter Optimization Framework}",
    author = "Akiba, Takuya and Sano, Shotaro and Yanase, Toshihiko and Ohta, Takeru and Koyama, Masanori",
    booktitle = "Proceedings of the 25th ACM SIGKDD International Conference on Knowledge Discovery and Data Mining",
    pages = "2623--2631",
    year = "2019"
}

@manual{DelphesHLLHCCard,
    title = "{Delphes 3 HL-LHC Card}",
    author = "Delphes Collaboration",
    year = "2022",
    url = "https://github.com/delphes/delphes/blob/master/cards/delphes_card_HLLHC.tcl",
    note = "GitHub repository: \url{https://github.com/delphes/delphes}"
}

@article{Alloul:2013bka,
    author = "Alloul, Adam and Christensen, Neil D. and Degrande, C{\'e}dric and Duhr, Claude and Fuks, Benjamin",
    title = "{FeynRules 2.0 - A complete toolbox for tree-level phenomenology}",
    journal = "Comput. Phys. Commun.",
    volume = "185",
    year = "2014",
    pages = "2250-2300",
    doi = "10.1016/j.cpc.2014.04.012",
    eprint = "1310.1921",
    archivePrefix = "arXiv",
    primaryClass = "hep-ph"
}

@article{Cerri:2018ypt,
    author = "Cerri, A. and others",
    editor = "Dainese, Andrea and Mangano, Michelangelo and Meyer, Andreas B. and Nisati, Aleandro and Salam, Gavin and Vesterinen, Mika Anton",
    title = "{Report from Working Group 4: Opportunities in Flavour Physics at the HL-LHC and HE-LHC}",
    eprint = "1812.07638",
    archivePrefix = "arXiv",
    primaryClass = "hep-ph",
    reportNumber = "CERN-LPCC-2018-06",
    doi = "10.23731/CYRM-2019-007.867",
    journal = "CERN Yellow Rep. Monogr.",
    volume = "7",
    pages = "867--1158",
    year = "2019"
}

@article{ATLAS:2023mvd,
    author = "Aad, Georges and others",
    collaboration = "ATLAS",
    title = "{Searches for lepton-flavour-violating decays of the Higgs boson into $e\tau$ and $\mu\tau$ in $\sqrt{s}=13$ TeV $pp$ collisions with the ATLAS detector}",
    eprint = "2302.05225",
    archivePrefix = "arXiv",
    primaryClass = "hep-ex",
    reportNumber = "CERN-EP-2022-279",
    doi = "10.1007/JHEP07(2023)166",
    journal = "JHEP",
    volume = "07",
    pages = "166",
    year = "2023"
}

@article{Arroyo-Urena:2020mgg,
    author = "Arroyo-Ure{\~n}a, M. A. and Valencia-P{\'e}rez, T. A. and Gait{\'a}n, R. and Montes De Oca, J. H. and Fern{\'a}ndez-T{\'e}llez, A.",
    title = "{Flavor-changing decay $h\to \tau\mu$ at super hadron colliders}",
    eprint = "2002.04120",
    archivePrefix = "arXiv",
    primaryClass = "hep-ph",
    doi = "10.1007/JHEP08(2020)170",
    journal = "JHEP",
    volume = "08",
    pages = "170",
    year = "2020"
}

@article{Urena:2021xtw,
    author = "Ure{\~n}a, Marco Antonio Arroyo and Gaitan-Lozano, Ricardo and de Oca Yemha, Jos{\'e} Halim Montes and V{\'e}lez, Ricardo S{\'a}nchez",
    title = "{Lepton flavor violating $h\rightarrow\tau\mu$ decay induced by leptoquarks}",
    doi = "10.31349/RevMexFis.67.040801",
    journal = "Rev. Mex. Fis.",
    volume = "67",
    number = "4",
    pages = "040801",
    year = "2021"
}

@article{Arroyo-Urena:2023vfh,
    author = "Arroyo-Ure{\~n}a, M. A. and D{\'\i}az-Cruz, J. Lorenzo and F{\'e}lix-Beltr{\'a}n, O. and Zeleny-Mora, M.",
    title = "{Lessons from LHC on the LFV Higgs decays $h\rightarrow \ell_a \ell_b$ in the two-Higgs doublet models}",
    eprint = "2308.01380",
    archivePrefix = "arXiv",
    primaryClass = "hep-ph",
    doi = "10.1142/S0217751X24500799",
    journal = "Int. J. Mod. Phys. A",
    volume = "39",
    number = "21",
    pages = "2450079",
    year = "2024"
}

@article{Lami:2016mjf,
    author = "Lami, Andrea and Roig, Pablo",
    title = "{$H\to \ell\ell'$ in the simplest little Higgs model}",
    eprint = "1603.09663",
    archivePrefix = "arXiv",
    primaryClass = "hep-ph",
    doi = "10.1103/PhysRevD.94.056001",
    journal = "Phys. Rev. D",
    volume = "94",
    number = "5",
    pages = "056001",
    year = "2016"
}

@article{Arganda:2004bz,
    author = "Arganda, Ernesto and Curiel, Ana M. and Herrero, Maria J. and Temes, David",
    title = "{Lepton flavor violating Higgs boson decays from massive seesaw neutrinos}",
    eprint = "hep-ph/0407302",
    archivePrefix = "arXiv",
    reportNumber = "FTUAM-04-16, IFT-UAM-CSIC-04-41, LAPTH-1057-04",
    doi = "10.1103/PhysRevD.71.035011",
    journal = "Phys. Rev. D",
    volume = "71",
    pages = "035011",
    year = "2005"
}

@article{Assamagan:2002kf,
    author = "Assamagan, Ketevi Adikle and Deandrea, Aldo and Delsart, Pierre-Antoine",
    title = "{Search for the lepton flavor violating decay $A^0 / H^0 \rightarrow \tau^\pm \mu^\mp$ at hadron colliders}",
    eprint = "hep-ph/0207302",
    archivePrefix = "arXiv",
    reportNumber = "LYCEN-2002-27, BNL-69300, ATL-COM-PHYS-2002-031",
    doi = "10.1103/PhysRevD.67.035001",
    journal = "Phys. Rev. D",
    volume = "67",
    pages = "035001",
    year = "2003"
}

@article{Han:2000jz,
    author = "Han, Tao and Marfatia, Danny",
    title = "{$h\rightarrow \mu\tau$ at hadron colliders}",
    eprint = "hep-ph/0008141",
    archivePrefix = "arXiv",
    reportNumber = "MADPH-00-1188",
    doi = "10.1103/PhysRevLett.86.1442",
    journal = "Phys. Rev. Lett.",
    volume = "86",
    pages = "1442--1445",
    year = "2001"
}

@article{Diaz-Cruz:1999sns,
    author = "Diaz-Cruz, J. Lorenzo and Toscano, J. J.",
    title = "{Lepton flavor violating decays of Higgs bosons beyond the standard model}",
    eprint = "hep-ph/9910233",
    archivePrefix = "arXiv",
    reportNumber = "HEP-IFUAP-99-2",
    doi = "10.1103/PhysRevD.62.116005",
    journal = "Phys. Rev. D",
    volume = "62",
    pages = "116005",
    year = "2000"
}

@article{Korner:1992zk,
    author = "Korner, J. G. and Pilaftsis, A. and Schilcher, K.",
    title = "{Leptonic CP asymmetries in flavor changing $H^0$ decays}",
    eprint = "hep-ph/9301289",
    archivePrefix = "arXiv",
    reportNumber = "MZ-TH-92-30",
    doi = "10.1103/PhysRevD.47.1080",
    journal = "Phys. Rev. D",
    volume = "47",
    pages = "1080--1086",
    year = "1993"
}

@article{Pilaftsis:1992st,
    author = "Pilaftsis, Apostolos",
    title = "{Lepton flavor nonconservation in $H^0$ decays}",
    reportNumber = "MZ-TH-92-10",
    doi = "10.1016/0370-2693(92)91301-O",
    journal = "Phys. Lett. B",
    volume = "285",
    pages = "68--74",
    year = "1992"
}

@article{BaBar:2009hkt,
    author = "Aubert, Bernard and others",
    collaboration = "BaBar",
    title = "{Searches for Lepton Flavor Violation in the Decays $\tau^\pm \to e^\pm \gamma$ and $\tau^\pm \to \mu^\pm \gamma$}",
    eprint = "0908.2381",
    archivePrefix = "arXiv",
    primaryClass = "hep-ex",
    reportNumber = "SLAC-PUB-13753, BABAR-PUB-09-026",
    doi = "10.1103/PhysRevLett.104.021802",
    journal = "Phys. Rev. Lett.",
    volume = "104",
    pages = "021802",
    year = "2010"
}

@article{MEG:2016leq,
    author = "Baldini, A. M. and others",
    collaboration = "MEG",
    title = "{Search for the lepton flavour violating decay $\mu^+ \rightarrow \mathrm{e}^+ \gamma$ with the full dataset of the MEG experiment}",
    eprint = "1605.05081",
    archivePrefix = "arXiv",
    primaryClass = "hep-ex",
    doi = "10.1140/epjc/s10052-016-4271-x",
    journal = "Eur. Phys. J. C",
    volume = "76",
    number = "8",
    pages = "434",
    year = "2016"
}

@article{SINDRUM:1987nra,
    author = "Bellgardt, U. and others",
    collaboration = "SINDRUM",
    title = "{Search for the Decay $\mu^+ \to e^+ e^+ e^-$}",
    reportNumber = "SIN-PR-87-09",
    doi = "10.1016/0550-3213(88)90462-2",
    journal = "Nucl. Phys. B",
    volume = "299",
    pages = "1--6",
    year = "1988"
}

@article{Hayasaka:2010np,
    author = "Hayasaka, K. and others",
    title = "{Search for Lepton Flavor Violating Tau Decays into Three Leptons with 719 Million Produced $\tau^+\tau^-$ Pairs}",
    eprint = "1001.3221",
    archivePrefix = "arXiv",
    primaryClass = "hep-ex",
    doi = "10.1016/j.physletb.2010.03.037",
    journal = "Phys. Lett. B",
    volume = "687",
    pages = "139--143",
    year = "2010"
}

@article{Maki:1962mu,
    author = "Maki, Ziro and Nakagawa, Masami and Sakata, Shoichi",
    title = "{Remarks on the unified model of elementary particles}",
    doi = "10.1143/PTP.28.870",
    journal = "Prog. Theor. Phys.",
    volume = "28",
    pages = "870--880",
    year = "1962"
}

@article{Pontecorvo:1957cp,
    author = "Pontecorvo, B.",
    title = "{Mesonium and Antimesonium}",
    journal = "Sov. Phys. JETP",
    volume = "6",
    pages = "429--431",
    year = "1958"
}

@article{K2K:2006yov,
    author = "Ahn, M. H. and others",
    collaboration = "K2K",
    title = "{Measurement of Neutrino Oscillation by the K2K Experiment}",
    eprint = "hep-ex/0606032",
    archivePrefix = "arXiv",
    doi = "10.1103/PhysRevD.74.072003",
    journal = "Phys. Rev. D",
    volume = "74",
    pages = "072003",
    year = "2006"
}

@article{Super-Kamiokande:2004orf,
    author = "Ashie, Y. and others",
    collaboration = "Super-Kamiokande",
    title = "{Evidence for an oscillatory signature in atmospheric neutrino oscillation}",
    eprint = "hep-ex/0404034",
    archivePrefix = "arXiv",
    doi = "10.1103/PhysRevLett.93.101801",
    journal = "Phys. Rev. Lett.",
    volume = "93",
    pages = "101801",
    year = "2004"
}

@article{KamLAND:2004mhv,
    author = "Araki, T. and others",
    collaboration = "KamLAND",
    title = "{Measurement of neutrino oscillation with KamLAND: Evidence of spectral distortion}",
    eprint = "hep-ex/0406035",
    archivePrefix = "arXiv",
    doi = "10.1103/PhysRevLett.94.081801",
    journal = "Phys. Rev. Lett.",
    volume = "94",
    pages = "081801",
    year = "2005"
}

@article{SNO:2002tuh,
    author = "Ahmad, Q. R. and others",
    collaboration = "SNO",
    title = "{Direct evidence for neutrino flavor transformation from neutral current interactions in the Sudbury Neutrino Observatory}",
    eprint = "nucl-ex/0204008",
    archivePrefix = "arXiv",
    doi = "10.1103/PhysRevLett.89.011301",
    journal = "Phys. Rev. Lett.",
    volume = "89",
    pages = "011301",
    year = "2002"
}

@article{Raidal:2008jk,
    author = "Raidal, M. and others",
    editor = "Fleischer, R. and Hurth, T. and Mangano, M. L.",
    title = "{Flavour physics of leptons and dipole moments}",
    eprint = "0801.1826",
    archivePrefix = "arXiv",
    primaryClass = "hep-ph",
    doi = "10.1140/epjc/s10052-008-0715-2",
    journal = "Eur. Phys. J. C",
    volume = "57",
    pages = "13--182",
    year = "2008"
}

@article{CMS:2021rsq,
    author = "Sirunyan, Albert M and others",
    collaboration = "CMS",
    title = "{Search for lepton-flavor violating decays of the Higgs boson in the $\mu\tau$ and $e\tau$ final states in proton-proton collisions at $\sqrt{s} = 13$ TeV}",
    eprint = "2105.03007",
    archivePrefix = "arXiv",
    primaryClass = "hep-ex",
    reportNumber = "CMS-HIG-20-009, CERN-EP-2021-061",
    doi = "10.1103/PhysRevD.104.032013",
    journal = "Phys. Rev. D",
    volume = "104",
    number = "3",
    pages = "032013",
    year = "2021"
}

@article{Arroyo-Urena:2025zvg,
    author = "Arroyo-Ure{\~n}a, M. A. and Carre{\~n}o, Diego and Valencia-P{\'e}rez, T. A.",
    title = "{Searching for the flavon in the diphoton channel at future super hadron colliders}",
    eprint = "2501.18675",
    archivePrefix = "arXiv",
    primaryClass = "hep-ph",
    doi = "10.1103/5btv-j8lv",
    journal = "Phys. Rev. D",
    volume = "111",
    number = "11",
    pages = "115037",
    year = "2025"
}

@article{Arroyo-Urena:2022oft,
    author = "Arroyo-Ure{\~n}a, Marco A. and Chakraborty, Amit and D{\'\i}az-Cruz, J. Lorenzo and Ghosh, Dilip Kumar and Khan, Najimuddin and Moretti, Stefano",
    title = "{Flavon signatures at the HL-LHC}",
    eprint = "2205.12641",
    archivePrefix = "arXiv",
    primaryClass = "hep-ph",
    doi = "10.1103/PhysRevD.108.095026",
    journal = "Phys. Rev. D",
    volume = "108",
    number = "9",
    pages = "095026",
    year = "2023"
}

@article{Abbas:2024dfh,
    author = "Abbas, Gauhar and Alok, Ashutosh Kumar and Chundawat, Neetu Raj Singh and Khan, Najimuddin and Singh, Neelam",
    title = "{Finding flavons at colliders}",
    eprint = "2407.09255",
    archivePrefix = "arXiv",
    primaryClass = "hep-ph",
    doi = "10.1103/PhysRevD.110.115015",
    journal = "Phys. Rev. D",
    volume = "110",
    number = "11",
    pages = "115015",
    year = "2024"
}

@article{Arroyo-Urena:2018mvl,
    author = "Arroyo-Ure{\~n}a, M. A. and D{\'\i}az-Cruz, J. L. and Tavares-Velasco, G. and Bola{\~n}os, A. and Hern{\'a}ndez-Tom{\'e}, G.",
    title = "{Searching for lepton flavor violating flavon decays at hadron colliders}",
    eprint = "1801.00839",
    archivePrefix = "arXiv",
    primaryClass = "hep-ph",
    doi = "10.1103/PhysRevD.98.015008",
    journal = "Phys. Rev. D",
    volume = "98",
    number = "1",
    pages = "015008",
    year = "2018"
}

@article{Bauer:2016rxs,
    author = "Bauer, Martin and Schell, Torben and Plehn, Tilman",
    title = "{Hunting the Flavon}",
    eprint = "1603.06950",
    archivePrefix = "arXiv",
    primaryClass = "hep-ph",
    doi = "10.1103/PhysRevD.94.056003",
    journal = "Phys. Rev. D",
    volume = "94",
    number = "5",
    pages = "056003",
    year = "2016"
}

@article{Arroyo-Urena:2019fyd,
    author = "Arroyo-Ure{\~n}a, M. A. and Fern{\'a}ndez-T{\'e}llez, A. and Tavares-Velasco, G.",
    title = "{Flavor changing flavon decay $\phi\to tc$ ($\phi = H_F, A_F$) at the high luminosity large hadron collider}",
    eprint = "1906.07821",
    archivePrefix = "arXiv",
    primaryClass = "hep-ph",
    doi = "10.31349/RevMexFis.69.020803",
    journal = "Rev. Mex. Fis.",
    volume = "69",
    number = "2",
    pages = "020803",
    year = "2023"
}

@article{Greljo:2024evt, 
    author = "Greljo, Admir and Smolkovi{\v{c}}, Aleks and Valenti, Alessandro",
    title = "{Froggatt-Nielsen ALP}",
    eprint = "2407.02998",
    archivePrefix = "arXiv",
    primaryClass = "hep-ph",
    doi = "10.1007/JHEP09(2024)174",
    journal = "JHEP",
    volume = "09",
    pages = "174",
    year = "2024"
}

@article{Loisa:2024xuk, 
    author = "Loisa, Eetu and Talbert, Jim",
    title = "{Froggatt-Nielsen meets the SMEFT}",
    eprint = "2402.16940",
    archivePrefix = "arXiv",
    primaryClass = "hep-ph",
    reportNumber = "LA-UR-24-21703",
    doi = "10.1007/JHEP10(2024)017",
    journal = "JHEP",
    volume = "10",
    pages = "017",
    year = "2024"
}

@article{Kikuchi:2023fpl, 
    author = "Kikuchi, Shota and Kobayashi, Tatsuo and Nasu, Kaito",
    title = "{CP phase in modular flavor models and discrete Froggatt-Nielsen models}",
    eprint = "2312.11809",
    archivePrefix = "arXiv",
    primaryClass = "hep-ph",
    doi = "10.1103/PhysRevD.109.115018",
    journal = "Phys. Rev. D",
    volume = "109",
    number = "11",
    pages = "115018",
    year = "2024"
}

@article{Asadi:2023ucx,
    author = "Asadi, Pouya and Bhattacharya, Arindam and Fraser, Katherine and Homiller, Samuel and Parikh, Aditya",
    title = "{Wrinkles in the Froggatt-Nielsen mechanism and flavorful new physics}",
    eprint = "2308.01340",
    archivePrefix = "arXiv",
    primaryClass = "hep-ph",
    doi = "10.1007/JHEP10(2023)069",
    journal = "JHEP",
    volume = "10",
    pages = "069",
    year = "2023"
}

@article{Alwall:2014hca,
    author = "Alwall, J. and Frederix, R. and Frixione, S. and Hirschi, V. and Maltoni, F. and Mattelaer, O. and Shao, H. -S. and Stelzer, T. and Torrielli, P. and Zaro, M.",
    title = "{The automated computation of tree-level and next-to-leading order differential cross sections, and their matching to parton shower simulations}",
    eprint = "1405.0301",
    archivePrefix = "arXiv",
    primaryClass = "hep-ph",
    reportNumber = "CERN-PH-TH-2014-064, CP3-14-18, LPN14-066, MCNET-14-09, ZU-TH-14-14",
    doi = "10.1007/JHEP07(2014)079",
    journal = "JHEP",
    volume = "07",
    pages = "079",
    year = "2014"
}

@article{Ball:2012cx,
    author = "Ball, Richard D. and others",
    title = "{Parton distributions with LHC data}",
    eprint = "1207.1303",
    archivePrefix = "arXiv",
    primaryClass = "hep-ph",
    reportNumber = "EDINBURGH-2012-08, IFUM-FT-997, FR-PHENO-2012-014, RWTH-TTK-12-25, CERN-PH-TH-2012-037, SFB-CPP-12-47",
    doi = "10.1016/j.nuclphysb.2012.10.003",
    journal = "Nucl. Phys. B",
    volume = "867",
    pages = "244--289",
    year = "2013"
}

@article{Sjostrand:2014zea,
    author = {Sj{\"o}strand, Torbj{\"o}rn and Ask, Stefan and Christiansen, Jesper R. and Corke, Richard and Desai, Nishita and Ilten, Philip and Mrenna, Stephen and Prestel, Stefan and Rasmussen, Christine O. and Skands, Peter Z.},
    title = "{An introduction to PYTHIA 8.2}",
    eprint = "1410.3012",
    archivePrefix = "arXiv",
    primaryClass = "hep-ph",
    reportNumber = "LU-TP-14-36, MCNET-14-22, CERN-PH-TH-2014-190, FERMILAB-PUB-14-316-CD, DESY-14-178, SLAC-PUB-16122",
    doi = "10.1016/j.cpc.2015.01.024",
    journal = "Comput. Phys. Commun.",
    volume = "191",
    pages = "159--177",
    year = "2015"
}

@article{deFavereau:2013fsa,
    author = "de Favereau, J. and Delaere, C. and Demin, P. and Giammanco, A. and Lema{\^\i}tre, V. and Mertens, A. and Selvaggi, M.",
    collaboration = "DELPHES 3",
    title = "{DELPHES 3, A modular framework for fast simulation of a generic collider experiment}",
    eprint = "1307.6346",
    archivePrefix = "arXiv",
    primaryClass = "hep-ex",
    doi = "10.1007/JHEP02(2014)057",
    journal = "JHEP",
    volume = "02",
    pages = "057",
    year = "2014"
}

@article{Harnik:2012pb,
    author = "Harnik, Roni and Kopp, Joachim and Zupan, Jure",
    title = "{Flavor Violating Higgs Decays}",
    eprint = "1209.1397",
    archivePrefix = "arXiv",
    primaryClass = "hep-ph",
    reportNumber = "FERMILAB-PUB-12-498-T",
    doi = "10.1007/JHEP03(2013)026",
    journal = "JHEP",
    volume = "03",
    pages = "026",
    year = "2013"
}

@article{Froggatt:1978nt,
    author = "Froggatt, C. D. and Nielsen, Holger Bech",
    title = "{Hierarchy of Quark Masses, Cabibbo Angles and CP Violation}",
    reportNumber = "CERN-TH-2519",
    doi = "10.1016/0550-3213(79)90316-X",
    journal = "Nucl. Phys. B",
    volume = "147",
    pages = "277--298",
    year = "1979"
}

@article{Dery_2014,
    title = "{Model building for flavor changing Higgs couplings}",
    volume = "90",
    ISSN = "1550-2368",
    url = "http://dx.doi.org/10.1103/PhysRevD.90.115022",
    DOI = "10.1103/physrevd.90.115022",
    number = "11",
    journal = "Physical Review D",
    publisher = "American Physical Society (APS)",
    author = "Dery, Avital and Efrati, Aielet and Nir, Yosef and Soreq, Yotam and Susi\v{c}, Vasja",
    year = "2014",
    month = dec
}

@article{Crivellin_2014,
    author = "Crivellin, Andreas and Hoferichter, Martin and Procura, Massimiliano",
    title = "{Improved predictions for $\mu\to e$ conversion in nuclei and Higgs-induced lepton flavor violation}",
    eprint = "1404.7134",
    archivePrefix = "arXiv",
    primaryClass = "hep-ph",
    reportNumber = "CERN-PH-TH-2014-068",
    doi = "10.1103/PhysRevD.89.093024",
    journal = "Phys. Rev. D",
    volume = "89",
    pages = "093024",
    year = "2014"
}

@article{Barman_2023,
    title = "{Constraining lepton flavor violating Higgs couplings at the HL-LHC in the vector boson fusion channel}",
    volume = "107",
    ISSN = "2470-0029",
    url = "http://dx.doi.org/10.1103/PhysRevD.107.075018",
    DOI = "10.1103/physrevd.107.075018",
    number = "7",
    journal = "Physical Review D",
    publisher = "American Physical Society (APS)",
    author = "Barman, Rahool Kumar and Dev, P. S. Bhupal and Thapa, Anil",
    year = "2023",
    month = apr
}

@article{Arganda_2015,
    title = "{Imprints of massive inverse seesaw model neutrinos in lepton flavor violating Higgs boson decays}",
    volume = "91",
    ISSN = "1550-2368",
    url = "http://dx.doi.org/10.1103/PhysRevD.91.015001",
    DOI = "10.1103/physrevd.91.015001",
    number = "1",
    journal = "Physical Review D",
    publisher = "American Physical Society (APS)",
    author = "Arganda, E. and Herrero, M. J. and Marcano, X. and Weiland, C.",
    year = "2015",
    month = jan
}

@article{Arganda_2016,
    title = "{Enhancement of the lepton flavor violating Higgs boson decay rates from SUSY loops in the inverse seesaw model}",
    volume = "93",
    ISSN = "2470-0029",
    url = "http://dx.doi.org/10.1103/PhysRevD.93.055010",
    DOI = "10.1103/physrevd.93.055010",
    number = "5",
    journal = "Physical Review D",
    publisher = "American Physical Society (APS)",
    author = "Arganda, E. and Herrero, M. J. and Marcano, X. and Weiland, C.",
    year = "2016",
    month = mar
}

@article{Arganda_2017,
    title = "{Effective lepton flavor violating vertex from right-handed neutrinos within the mass insertion approximation}",
    volume = "95",
    ISSN = "2470-0029",
    url = "http://dx.doi.org/10.1103/PhysRevD.95.095029",
    DOI = "10.1103/physrevd.95.095029",
    number = "9",
    journal = "Physical Review D",
    publisher = "American Physical Society (APS)",
    author = "Arganda, E. and Herrero, M. J. and Marcano, X. and Morales, R. and Szynkman, A.",
    year = "2017",
    month = may
}

@article{Marcano_2020,
    title = "{Flavor Techniques for LFV Processes: Higgs Decays in a General Seesaw Model}",
    volume = "7",
    ISSN = "2296-424X",
    url = "http://dx.doi.org/10.3389/fphy.2019.00228",
    DOI = "10.3389/fphy.2019.00228",
    journal = "Frontiers in Physics",
    publisher = "Frontiers Media SA",
    author = "Marcano, Xabier and Morales, Roberto A.",
    year = "2020",
    month = jan
}

@article{Asadi_2026,
    title = "{Lepton flavor violation: From muon decays to muon colliders}",
    volume = "113",
    ISSN = "2470-0029",
    url = "http://dx.doi.org/10.1103/bg4z-dmgb",
    DOI = "10.1103/bg4z-dmgb",
    number = "1",
    journal = "Physical Review D",
    publisher = "American Physical Society (APS)",
    author = "Asadi, Pouya and Bagherian, Hengameh and Fraser, Katherine and Homiller, Samuel and Lu, Qianshu",
    year = "2026",
    month = jan
}

@article{Crivellin:2015mga,
    author = "Crivellin, Andreas and D'Ambrosio, Giancarlo and Heeck, Julian",
    title = "{Explaining $h\to\mu^\pm\tau^\mp$, $B\to K^* \mu^+\mu^-$ and $B\to K \mu^+\mu^-/B\to K e^+e^-$ in a two-Higgs-doublet model with gauged $L_\mu-L_\tau$}",
    eprint = "1501.00993",
    archivePrefix = "arXiv",
    primaryClass = "hep-ph",
    reportNumber = "CERN-PH-TH-2015-001, ULB-TH-14-26",
    doi = "10.1103/PhysRevLett.114.151801",
    journal = "Phys. Rev. Lett.",
    volume = "114",
    pages = "151801",
    year = "2015"
}

@inproceedings{Chen:2016:XST:2939672.2939785,  
    author = {Chen, Tianqi and Guestrin, Carlos},  
    title = "{XGBoost: A Scalable Tree Boosting System}",  
    booktitle = {Proc. 22nd ACM SIGKDD Int. Conf. on Knowledge Discovery and Data Mining},  
    series = {KDD '16},  
    year = {2016},  
    isbn = {978-1-4503-4232-2},  
    location = {San Francisco, California, USA},  
    pages = {785},  
    numpages = {10},  
    doi = {10.1145/2939672.2939785},  
    acmid = {2939785},  
    publisher = {ACM},  
    address = {New York, NY, USA}
}

@article{ATLAS:2021vrm,
    collaboration = "ATLAS",
    title = "{Combined measurements of Higgs boson production and decay using up to $139$ fb$^{-1}$ of proton-proton collision data at $\sqrt{s}= 13$ TeV collected with the ATLAS experiment}",
    reportNumber = "ATLAS-CONF-2021-053",
    year = "2021"
}

\end{document}